\documentclass[11pt,authoryear]{elsarticle}

\usepackage{microtype}

\usepackage{setspace}

\usepackage{enumitem} 
\usepackage{amsthm}
\usepackage{amssymb} 

\usepackage{float} 

\usepackage{threeparttable}

\usepackage{graphicx}

\usepackage{subcaption} 

\usepackage{mathtools}

\usepackage{comment}
 
 \usepackage[table]{xcolor} 

 \usepackage{bm}
\usepackage{latexsym}
\usepackage{mathtools}

 \usepackage{tikz}
\usepackage{tikz-cd}
\usetikzlibrary{shapes.geometric, arrows.meta, positioning, fit, shadows.blur, calc}
\tikzcdset{diagrams={nodes={inner sep=7.5pt}}}

\usepackage{pgfplots}
\pgfplotsset{width=12cm, compat=1.18} 

\usepackage{enumitem} 
\setlist{noitemsep, topsep=0pt, parsep=0pt, partopsep=0pt}

\pgfplotsset{every axis/.append style={line width=0.6pt, tick style={line width=0.5pt}}}

\usepackage{booktabs}
\usepackage{multirow}
\usepackage{array}
\usepackage{tabularx}
\newcolumntype{M}[1]{>{\centering\arraybackslash}m{#1}}

\usepackage[noend]{algorithm2e}
\SetKwFor{ParallelFor}{ParFor}{do in parallel}{end}
\SetKwFor{Parallel}{}{Do in parallel}{end}
\RestyleAlgo{ruled}
\SetKwComment{Comment}{//* }{ *//}

\newtheorem{theorem}{Theorem}
\newtheorem{lemma}[theorem]{Lemma}

\newtheorem{definition}{Definition}

\theoremstyle{definition} 

\numberwithin{equation}{section}

\usepackage[all]{xy}
\usepackage{comment}
\usepackage{caption}
\usepackage{url}
\usepackage{hyperref}

\newcommand{\BibTeX}{\rm B\kern-.05em{\sc i\kern-.025em b}\kern-.08em\TeX}
\def\permille{\ensuremath{{}^\text{o}\mkern-5mu/\mkern-3mu_\text{oo}}}

\begin{document}
 
 \begin{frontmatter}

\title{Robust large-scale approximate solutions of the constrained Kemeny aggregation ranking problem}

\author[inst1]{Xuan Kien Phung}
\ead{xuan.kien.phung@umontreal.ca}  
\author[inst1]{Sylvie Hamel}
\ead{sylvie.hamel@umontreal.ca}

\affiliation[inst1]{organization={Département d'informatique et de recherche opérationnelle, Université de Montréal},
            city={Montréal},
            postcode={H3T 1J4}, 
            state={QC},
            country={Canada}}

\begin{abstract}
The Kemeny aggregation ranking problem consists of computing the consensus rankings of an election with respect to the well-known Kemeny-Young voting method. These consensus rankings satisfy many fundamental properties and are the geometric medians of the votes in the election under the Kendall-tau distance which counts the number of pairwise disagreements. While having various applications in computational social choice, machine learning, operations research, and biology, the problem of computing the consensus rankings is NP-hard and encounters a severe bottleneck in practice. In this paper, we consider the more general Kemeny aggregation ranking problem under immutable input pairwise constraints and propose robust and scalable approximate methods with provable guarantees.  We thus provide an algorithmic solution to problem of incorporating a posteriori immutable pairwise constraints into past rankings (e.g., historical voter ballots or legacy laboratory assays) without incurring the cost of generating new data.
Our base algorithms $\pmb \Gamma$-\textbf{XCC}, \textbf{\textup{M-INCR-DEG}}, \textbf{\textup{{LP}-M-INCR-DEG}}, and  \textbf{\textup{LP}}-$\pmb {\Gamma}$-\textbf{\textup{XCC}} have polynomial time complexity and  provides a source of rankings with bounded constant approximation factors.  Notably, $\pmb \Gamma$-\textbf{XCC} has quadratic runtimes, extends the Copeland ranking method, and satisfies the constrained Condorcet consistency property, an extension of the Condorcet winner criterion for elections with immutable pairwise constraints.  \textbf{\textup{{LP}-M-INCR-DEG}} and  \textbf{\textup{LP}}-$\pmb {\Gamma}$-\textbf{\textup{XCC}} provide feasible solutions respecting the input constraints and when combined with the quadratic time \textbf{M-XCC} partitioning, they also verify the constrained Condorcet consistency property. 
Thereby, our methods address the reliability and explainability issues. Our algorithms are applicable to weighted tournaments  with input pairwise constraints and satisfy similar properties in this larger context.  
\end{abstract}


\end{frontmatter}

\section{Introduction} 
The Kendall-tau distance between two rankings counts the minimal number of swaps of adjacent candidates to transform one ranking to another. The Kemeny aggregation ranking problem for elections \cite{kemeny, kemeny-snell,young,kendall-tau}  asks to compute an optimal consensus ranking called median of a given election over a set of $n$ candidates.  A ranking is optimal if it minimizes the total  Kendall-tau distance to the votes in the voting profile. A  median is also a maximum likelihood estimator of the central ranking  \cite{kemeny} in the statistical Mallows model \cite{mallows}. The Kemeny voting rule specifies that the result of an election is the set of  medians.  
In social choice theory, the Kemeny voting rule  is the unique rule with the majority, neutrality, consistency, and Condorcet winner properties   \cite{young-levenglick}. For this reason, it has wide  application  in computational social choice theory  and decision theory \cite{recommender, social, young, young-levenglick}, in bioinformatics  \cite{andrieu, brancotte, hamel-bio, 3/4}, in recommender systems \cite{Recommender-1}, and in 
machine learning and artificial intelligence \cite{3/4,davenport, Kemenywww, soft-condorcet}. Unfortunately, computing a median  is   an NP-hard problem \cite{dwork, bachmeier}. Devising methods to obtain large-scale and provable high quality approximate solutions constitutes a reasonable approach in real world application. 
\par 
There is a 
notorious challenge in the empirical sciences, notably in gene regulatory
network inference and protein pathway mapping. In these domains, input
rankings originate from diverse high-throughput laboratory assays that
are intrinsically corrupted by  instrumental noise and experimental
artifacts. While minimizing this noise is critical, a major challenge arises
when immutable, domain-specific expert knowledge, such as strict metabolic
sequence constraints, is discovered independently and a posteriori. Generating empirical genomic profiles represents an immense financial burden
and a highly energy-intensive computational footprint.  Discarding legacy
datasets or re-executing physical assays is thus unviable from an
eco-efficiency standpoint. Consequently, there is a critical, unaddressed need
for a framework capable of retroactively salvaging and rectifying historical,
noisy ranking profiles by injecting a late-stage set of ground-truth constraints
provided by experts. 
\par 
\emph{Previous work}. Notable existing approximate algorithms of the Kemeny aggregation ranking problem include Spearmans' footrule \cite{spearman-1904, spearman-1906},  the Copeland method \cite{copeland}, the \textbf{INCR-INDEG} algorithm   \cite{winners-ordering}, and the Approximate Condorcet Partitioning (\textbf{ACP}) \cite{escobedo}. Spearman’s footrule distance evaluates the distance between two rankings by accumulating the mismatches and summing the distances of the positions of each candidate. The ranking minimizing the Spearman’s footrule distance can be computed in $O(n\ln n)$ and provides a bounded 2-approximation factor for the Kemeny aggregation ranking problem. Here, $n=|C|$ and $C$ is the set of candidates in the problem. 
The Copeland algorithm simply ranks the candidates by their number of pairwise  wins against other candidates, where ties are broken arbitrarily.  
 The Copeland method provides a 4-approximation solution, resp. a 9-approximation solution,  to the Kemeny aggregation ranking problem for elections, resp. for weighted tournament. Moreover, the time complexity of the Copeland  algorithm is $O(n^2)$.  
 If we rank candidates increasingly according to their total in-degrees in the weighted tournament, we obtain the 5-approximation algorithm \textbf{INCR-INDEG} \cite{winners-ordering} of the Kemeny aggregation ranking problem for weighted tournaments.  Experimental results show that the rankings returned by the Copeland  algorithm  tend to have better quality (lower Kemeny scores) than rankings returned by \textbf{INCR-INDEG}, notably for instances where the Finest Condorcet Partitioning performs well. 
\cite{escobedo} introduces a scalable approximate solution \textbf{ACP} based on the extended Condorcet partition (\textbf{XCC}). It is an unsolved   problem in their paper whether such ACP solutions 
 {provide a constant-factor approximation for the Kemeny  aggregation ranking problem}. Among these algorithms, only the Copeland method satisfies the Condorcet consistency criterion while others may fail to rank a Condorcet winner first. 
\par 
\emph{Our work}. The first main contribution of this paper concerns  robust and scalable approximate methods for the Kemeny aggregation ranking problem over weighted tournaments with immutable input pairwise constraints. 
 For this,  we introduce the general algorithm $\bm \Gamma$-\textbf{XCC} 
  and study its theoretical properties. Our algorithm $\bm \Gamma$-\textbf{XCC} generalizes the  Copeland method and satisfies the generalized Condorcet consistency property for constrained weighted  tournaments. We also prove that  $\bm \Gamma$-\textbf{XCC} is a 4-approximation, resp. a 9-approximation, of the Kemeny aggregation ranking problem for constrained elections, resp. for constrained weighted  tournaments.      Generalizing the \textbf{ACP} method, we introduce the \textbf{M-ACP} method in the context of weighted tournaments with input pairwise constraints.  As for \textbf{ACP} is an approximate version of the finest \textbf{XCC} partitioning technique, \textbf{M-ACP} is an approximate version of the finest \textbf{M-XCC} partitioning method introduced in \cite{phung-hamel-exact}.  To improve \textbf{INCR-DEG} in the context of constrained weighted tournaments, we introduce the \textbf{M-INCR-DEG} algorithm that takes into consideration the set of input constraints and still produces a 5-approximation of the constrained Kemeny aggregation ranking  problem. 
   \par 
  For our second contribution, we introduce the algorithms \textbf{\textup{{LP}-M-INCR-DEG}} and  \textbf{\textup{LP}}-$\pmb {\Gamma}$-\textbf{\textup{XCC}} whose ideas are similar to  $\bm \Gamma$-\textbf{XCC}  and   \textbf{M-INCR-DEG} but they further exploit the optimal LP solution of the constrained Kemeny aggregation ranking problem. Critically, the output rankings of these algorithms are feasible solutions in the sense that they satisfy all the input pairwise constraints of the problem.  \textbf{\textup{{LP}-M-INCR-DEG}} and  \textbf{\textup{LP}}-$\pmb {\Gamma}$-\textbf{\textup{XCC}}  are respectively 4-approximation and 8-approximation of the Kemeny aggregation ranking  problem for constrained elections. By applying \textbf{\textup{{LP}-M-INCR-DEG}} and  \textbf{\textup{LP}}-$\pmb {\Gamma}$-\textbf{\textup{XCC}}  to each block in the $M$-XCC partition where $M\subset C^2 $ is the set of input constraints, the resulting rankings are not only feasible solutions with bounded constant-approximation facotrs but also satisfy the constrained Condorcet property.    
  \par 
  Our third contribution concerns a pairwise constraint discovery engine called $\alpha$MOT* in the context of the Kemeny aggregation ranking problem over weighted tournaments with immutable input pairwise constraints. We obtain $\alpha$MOT* as a modified version of the $\alpha$MOT method in \cite{phung-hamel-exact}. The pairwise constraints discovered by $\alpha$MOT may violate the set of initial immutable input constraints notably  when this input set is not satisfied by some optimal consensus ranking of the unconstrained problem.  In this aspect, the constraints discovered by $\alpha$MOT* are guaranteed to be satisfied by the optimal rankings for the constrained problem. Similarly, we also introduce $\alpha$MOTe* as a modified version of the method $\alpha$MOTe obtained in \cite{phung-hamel-exact} to deal with immutable input pairwise constraints of the problem. 
\par 
Our fourth contribution is a greedy version G1 of $\alpha$MOTe*.  While it is not guaranteed that some median can satisfy the whole set of constraints obtained by G1, we obtain a significantly larger set of constraints than that of  $\alpha$MOTe. Moreover, M-ACP approximate solutions built from G1 achieve very good approximation factors, meaning most of G1 constraints are satisfied by some optimal solution of the problem. 
 These M-ACP approximate solutions are in fact optimal for a significant   portion of instances tested. 
 \par 
To exploit the set of constraints $M$ obtained by $\alpha$MOT* or $\alpha$MOTe* or G1, we note that $M$ contains the initial input constraints of the problem. We can feed $M$ to the decomposition algorithm \textbf{M-XCC} in to compute the finest $M$-XCC partition applicable to all optimal rankings satisfying $M$. We can also leverage such set of constraints $M$ to reduce the size of the IP model of the constrained problem and accelerate the runtime of the IP solvers as discussed in \cite{phung-hamel-exact}. 
\section{Backgrounds and notations}
\label{s:background}

 Following the terminology in social choice theory, an election $(C,V)$ is a finite collection $C$ of $n=|C|$ candidates (or alternatives) together with a voting profile $V$ consisting of $m$ votes with repetitions allowed. In this paper, a vote (or ranking)  is simply a complete and strict total ordering presented as a bijection $\pi \colon C \to \{0,1,\dots, n-1\}$ where $\pi(x)$ is the position of $x$ in the ranking $\pi$. We use the notations $x>^\pi y$ and  $\pi(x)<\pi(y)$ interchangeably which mean $x$ is ranked before $y$ (or $x>y$) in $\pi$.   
Let $S(C)$ be the set of all rankings over the set of candidates $C$. 
The \emph{Kendall-tau distance} $ d_{KT}(\pi, \sigma)$ between two rankings $\pi, \sigma$ of $C$ is: 
\begin{equation}
\label{e:definition-k-wise-distance}
    d_{KT}(\pi, \sigma) = \sum_{\{x,y\} \subset C} \textbf{1}({\mathrm{top}(\pi,{\{x,y\}}) \neq  \mathrm{top}(\sigma,{\{x,y\}})}) \vspace{-0.2cm}
\end{equation}
where $\mathrm{top}(\pi,{\{x,y\}}) \in \{x,y\}$ denotes the highest ranked element in the induced ranking $\pi\vert_{\{x,y\}}$ of $\pi$ on $\{x,y\}$ and the indicator function $\mathbf{1}(X)=1$ if $X$ is true and $\mathbf{1}(X)=0$ if $X$ is false. 
The Kendall-tau distance between a ranking $\pi$ of $C$ and a collection of rankings $A$ of $C$ is defined formally as $
    d_{KT}(\pi,A) = \displaystyle \sum_{\sigma \in A}  d_{KT}(\pi,\sigma)$. A ranking $\pi \in S(C)$ is a \emph{median} of $(C,V)$ if 
$   d_{KT}(\pi,V) = \displaystyle \min_{\sigma \in S(C)} d_{KT} (\sigma, V)$. 
We can associate a weighted majority graph to $(C,V)$ where the set of vertices is $C$ and an arc from a vertex $x$ to another vertex $y$ is weighted by the proportion $a_{xy}$ of the votes in which $x$ is ranked before $y$. 
Equivalently, this weighted majority graph can be presented by the matrix  $[\delta_{xy}]_{x,y \in C}$ where 
$\delta_{xy} =  a_{xy} - a_{yx}$  is the net preference proportion.  
\par 
In this paper, we consider more generally the {Kemeny aggregation problem with  immutable input pairwise constraints}. More specifically, given    a set of pairwise constraints $W\subseteq C^2$, let  $(C,V,W)$ denote the corresponding constrained election. We address the constrained optimization problem 
\begin{align}
\label{e:general}
  \min_{\sigma \in S(C)} \quad & d_{KT} (\sigma, V)  \\
\mathrm{s.t.}\quad  
&  x > y \text{ in } \sigma \text{ for all } (x,y) \in W.  \nonumber  
\end{align}
 Every optimal solution of this general problem is called a median of the constrained election $(C,V,W)$. Such a  median is simple an optimal ranking among all rankings respecting the constraints $W$, and is in general not a median of the unconstrained weighted election $(C,V)$, unless some median of $(C,V)$ satisfies all the constraints in $W$. We observe that the computing difficulty increases significantly when no median of the unconstrained weighted election $(C,V)$ can satisfy the whole set of constraints $W$.  
 \par 
The Kemeny aggregation ranking problem can be extended to the weighted tournament setting as follows. 
Consider a weighted tournament on $C$ whose weight matrix $[a_{xy}]_{x,y \in C}$   is only required to satisfy the probability constraints $a_{xy}+a_{yx}=1$ for all $x\neq y$ and $a_{xx}=0$ for all $x$.    We denote the normalized Kemeny score of a ranking $\pi$ with respect to the tournament by 
$$d_{KT}(\pi) = \sum_{\{x,y\}\subset C} ( \mathbf{1}(x>^\pi y) a_{yx} + \mathbf{1}(y>^\pi x) a_{xy}).$$ 
Every ranking $\pi^*$ that minimizes the normalized Kemeny score $d_{KT}(\pi)$ is called a median of the tournament. Similarly, when subject to a set of immutable pairwise constraints $W\subset C^2$, a  ranking $\pi^*$ that satisfies $W$  and minimizes $d_{KT}(\pi)$  is called a median of the   constrained tournament. 
\par 
The tournament becomes an election if there exists a voting profile $V$  which realizes the matrix $[a_{xy}]_{x,y \in C}$, that is,  $a_{xy}$ is the proportion of votes in $V$ that rank $x$ before $y$. In this case, we also say that the tournament is realizable and note that 
$d_{KT}(\pi) = |V| d_{KT}(\pi, V)$. 
\par 
Unless stated otherwise, every set $U\subset C^2$ of constraints in this paper must be non self-contradictory in the sense that it does not contains any cycle when $U$ is viewed as a directed graph. In other words, $U$ is  non self-contradictory if its transitivity closure is a partial order on $C$.

\section{$\alpha$MOT* for constrained weighted tournament}

Let $U\subseteq C^2$ be a set of constraints. The \emph{transitive closure} of $U$ (the smallest transitive subset of $C^2$ containing $U$)  is denoted by $\overline{U}  \subseteq C^2$.  
For $x\in C$, we define  
    $L_{x}(U)  =\{z \colon (z,x)  \in {U} \}$ and $
 R_{x}(U)  =\{z \colon (x,z) \in {U} \}$  as the set of  elements ranked to the left and to the right  of $x$ according to the set of constraints $U$, respectively. We recall the following pairwise constraint generator $\alpha$MOT introduced in \cite{phung-hamel-exact} for  constrained weighted tournament with valid input constraints.

\begin{theorem}[$\alpha$MOT]
\label{t:alpha-MOT} 
Consider a  constrained weighted tournament over  $C$   with valid constraints $W\subseteq C^2$.  
Let $W_{-1}=W$ and 
for $k \geq 0$, define $W_k$  inductively as the set pairs $(x,y)$ of distinct candidates such that  \begin{align}
\label{e:alpha-beta-2-mot-condition}
     \delta_{xy}  > \min_{\alpha, \beta\geq 0,\,  \alpha+\beta=1} \sum_{z\in\,  Z_k(x,y)}  \max\left(0, \alpha \delta_{yz} + \beta \delta_{zx} \right),
\end{align}
where $ 
Z_k(x,y)= C\setminus \left( L_{y}\left(\overline{W}_{k-1}\right) \cup R_{x}\left(\overline{W}_{k-1}\right) \cup \{x,y\}\right)$ 
is the interference set. 
Suppose that  $(x,y) \in \overline{W}_{k}$ for  some   $k \geq -1$. Then $x$ is ranked before $y$ in every median of the constrained weighted tournament.  
\end{theorem}
 
To deal with an arbitrary set of input pairwise constraints $W\subset C^2$ or when there is no  certification of the validity of the input constraints in $W$, we must strengthen the condition of the method $\alpha$MOT. The resulting method is called $\alpha$MOT*. 

\begin{theorem}[$\alpha$MOT$^*$]
 \label{t:alpha-MOT-star}  
Consider a constrained weighted tournament over  $C$   with constraints $W\subseteq C^2$.  
Let $W_{-1}=W$ and 
for $k \geq 0$, define $W_k$  inductively as the set pairs $(x,y)$ of distinct candidates such that at least one of the following conditions hold with $Z_k(x,y) = C\setminus \left( L_{y}\left(\overline{W}_{k-1}\right) \cup R_{x}\left(\overline{W}_{k-1}\right) \cup \{x,y\}\right)$: 
\begin{enumerate}[label=\arabic*.] 
\item 
$(z,x) , (y,z)\notin W$ for all $z\in Z_k(x,y)$  
and  \begin{align}
\label{e:alpha-beta-2-mot-condition}
     \delta_{xy}  > \min_{\alpha, \beta\geq 0,\,  \alpha+\beta=1} \sum_{z\in\,  Z_k(x,y)}  \max\left(0, \alpha \delta_{yz} + \beta \delta_{zx} \right)
\end{align}
\item $(z,x)\notin W$ for all $z\in Z_k(x,y)$  
and  
    $ \delta_{xy}  > \sum_{z\in\,  Z_k(x,y)}  \max\left(0, \delta_{zx} \right)$. 
\item $(y,z)\notin W$ for all $z\in Z_k(x,y)$  
and 
    $  \delta_{xy}  > \sum_{z\in\,  Z_k(x,y)}  \max\left(0, \delta_{yz} \right)$. 
\end{enumerate} 
Suppose that  $(x,y) \in \overline{W}_{k}$ for  some   $k \geq -1$. Then $x$ is ranked before $y$ in every median of the constrained election $(C,V,W)$.  
\end{theorem}

 \begin{proof} The proof  generalizes directly the proof of  $\alpha$MOT given in \cite{phung-hamel-exact}. The case $k=-1$ is trivial as $W_{-1}=W$. Suppose that the conclusion holds up to $k \geq -1$.  For case 1, suppose on the contrary that  there exists $(x,y) \in W_{k+1}$ with $(z,x) , (y,z)\notin W$ for all $z\in Z_k(x,y)$ and  a median $\pi \colon LyZxR$  where $L,Z,R$ are ordered sets. Let $\alpha, \beta \geq 0$ with $\alpha+\beta=1$ and $\sum_{z\in Z_k(x,y)}  \max\left(0, \alpha \delta_{yz} + \beta \delta_{zx} \right)< \delta_{xy}$. 
  By hypothesis, 
$L_{y}\left(\overline{W}_{k}\right) \subseteq L$ and $R_{x}\left(\overline{W}_k\right) \subseteq R$ so   
$ 
Z \subseteq Z_{k+1}(x,y) = C\setminus \left(L_{y}\left(\overline{W}_k\right) \cup R_{x}\left(\overline{W}_k\right) \cup \{x,y\} \right)$. 
Let  $\sigma_1 \colon LZxyR$,  $\sigma_2\colon LxyZR$, and $
 \Delta_i= d_{KT} (\pi, V) - d_{KT}(\sigma_i, V)$.  Since $(z,x) , (y,z)\notin W$ for all $z\in Z$, the rankings $\sigma_1, \sigma_2$ satisfy $W$. 
As  $Z\subseteq Z_{k+1}(x,y)$, we have    
$ \alpha \Delta_1 +  \beta \Delta_2  = \alpha\left(  \delta_{xy} -   \sum_{z \in Z}  \delta_{yz} \right) + \beta \left(  \delta_{xy} -   \sum_{z \in Z}\delta_{zx} \right)    = (\alpha+\beta) \delta_{xy} -    \sum_{z \in Z} \left(\alpha \delta_{yz} + \beta \delta_{zx} \right) =   \delta_{xy} -    \sum_{z \in Z} \left(\alpha  \delta_{yz} + \beta \delta_{zx} \right) 
\geq \delta_{xy} -    \sum_{z \in Z_{k+1}(x,y)} \max \left(0, \alpha \delta_{yz} + \beta \delta_{zx} \right) >0$. 
Hence, $\Delta_j >0$ for some $j \in \{1,2\}$ and $\sigma_j$ has lower Kemeny score  than $\pi$, which contradicts the choice of $\pi$. Thus, $x$ must be ranked before $y$ in every  median. By transitivity, we conclude that  $x$ is ranked before $y$ in every  median if $(x,y) \in \overline{W}_{k+1}$. The proofs of case 2 and case 3 are similar where we set $\beta=1$ and $\alpha=1$ respectively. 
\end{proof}

Algorithm~\ref{algo:abmot-star-single-pair} below describes the single iteration $\alpha$MOT* verification for a given pair $(x,y)$ and a given input set of pairwise constraints $W\subset C^2$. Algorithm~\ref{algo:abmot-star-single-pair} returns a variable RHS. The pair $(x,y)$ is a valid constraint according to Theorem~\ref{t:alpha-MOT-star} if RHS is a number and $\delta_{xy}>\mathrm{RHS}$.  By replacing the algorithm $\alpha$MOT-SinglePair in the paper \cite{phung-hamel-exact} by  Algorithm~\ref{algo:abmot-star-single-pair}, we obtain from their work an efficient parallel framework to compute the whole set of $\alpha$MOT* constraints. As for $\alpha$MOT-SinglePair, our constrained version $\alpha$MOT-SinglePair* also has quadratic time complexity $O(n^2)$ where $n=|C|$.  
Obviously, when the set immutable input pairwise constraints $W$ is valid constraints in the sense that $W$ is satisfied by some median of the associated unconstrained problem, then we simply invoke  $\alpha$MOT instead of the restrictive version  $\alpha$MOT*.

\begin{algorithm}[ht]
\caption{$\alpha$MOT-SinglePair*} 
\label{algo:abmot-star-single-pair}
\DontPrintSemicolon 
\KwData{candidates $x,y\in C$, matrix $[\delta_{uv}]_{u,v\in C}$, constraints $W\subseteq C^2$}
\KwResult{ $\displaystyle \mathrm{RHS} $} 
$Z \gets C\setminus \left( L_{y}\left({W}\right) \cup R_{x}\left({W} \right) \cup \{x,y\}\right)$ \; 
$u \gets \mathbf{1}((z,x)\in W \textup{ for some } z\in Z )$ \; 
$v \gets \mathbf{1}((y,z)\in W \textup{ for some } z\in Z )$ \;  
\If{$(u,v)=(1,1)$}{\KwRet{$\mathrm{False}$}}
\If{$(u,v)=(0,0)$}{\KwRet{$ \min_{\alpha, \beta\geq 0,\,  \alpha+\beta=1} \sum_{z\in Z}  \max\left(0, \alpha \delta_{yz} + \beta \delta_{zx} \right)$}}
\If{$(u,v)=(1,0)$}{\KwRet{$  \sum_{z\in Z}  \max\left(0, \delta_{yz}  \right)$}}
\If{$(u,v)=(0,1)$}{\KwRet{$  \sum_{z\in Z}  \max\left(0,  \delta_{zx} \right)$}} 
\end{algorithm}

 \section{$\alpha$MOTe*, LowInf, and the greedy extension G1}
 \label{s:aMOTe-G1-G2}

 We can relax the strict inequality in the condition of $ \alpha$MOT* to obtain more constraints at the price that several such equality-type constraints may altogether exclude all the medians and thus lead to suboptimal solutions.   
To solve this issue,  we propose the following solution to the second issue that allows us to add an extra set of valid constraints satisfied by \textit{some} median of the constrained weighted tournament.

 \begin{theorem}[$\alpha$MOT{e}*]
 \label{alg:abmote}
Let $E$ be the set of all ordered pairs $(x,y)$ with  $(z,x) , (y,z)\notin W$ for all $z\in Z_k(x,y)$ that realizes the equality in the  condition~\eqref{e:alpha-beta-2-mot-condition} for at least one value of $k$. Then for every  $S \subseteq C$, there exists some median of the constrained weighted tournament satisfying all the constraints in $E(S)=\{(x,y)\in E\colon x\in S, \, y \notin S\}$. 
\end{theorem}
The proof of  $\alpha$MOT{e}* follows the same lines in the proof of $\alpha$MOTe given in \cite{phung-hamel-exact}. 
Computing a subset $S\subseteq C$ maximizing $|E(S)|$ is an instance of the Directed Maximum Cut Problem. In this case and when the input $W$ for the algorithm $\alpha$MOT* is the initial immutable input pairwise constraints of the problem, then we call the resulting total set of constraints $\alpha$MOTe* constraints or simply $\alpha$MOTe*. 
\par 

To obtain even more constraints, we propose  Algorithm~\ref{alg:abMOTe-greedy} called $\alpha$MOTe-Greedy* as an approximate version of $\alpha$MOTe* where we try to accept as many pairs as possible with minimal consideration on the validity of the obtained set of constraints. 
For this, we take the transitive closure of the set of constraints every time a new constraint is found to avoid having two opposite and contradictory constraints $(x,y)$ and $(y,x)$.

 \begin{algorithm}[hbt!]
\caption{ $\alpha$MOTe-Greedy*}\label{alg:abMOTe*-greedy}
\label{alg:abMOTe-greedy}
\DontPrintSemicolon 
\KwData{candidates $C$, matrix $[\delta_{uv}]_{u,v\in C}$, constraints $W \subseteq C^2$}
\KwResult{a new set of constraints $M\subseteq C^2$ containing $W$}  
$\mathrm{len} \gets -1$ \; 
 \While{$\mathrm{len} < |W|$}{
  $\mathrm{len} \gets |W|$\;
  {\For{$(x,y) \in C^2$ \textup{with} $\delta_{xy}>0$, $(x,y)\notin W$}{$\mathrm{RHS} \gets \alpha\text{MOT-SinglePair*}(x,y,[\delta_{uv}]_{u,v\in C}, W)$\; 
   \If{$\mathrm{RHS}=\mathrm{False}$}{Do not append $(x,y)$ to W, consider next pair\;
    }
    \ElseIf{$\delta_{xy} \geq \mathrm{RHS}$}{append  $(x,y)$ to $W$\;
    $W \gets \overline{W}$\;
    }
    }
  }
} 
$M \gets W$\; 
\KwRet{$M$} 
\end{algorithm}

If the input $W\subset C^2$ of $\alpha$MOTe-Greedy* is the set of constraints $\alpha$MOTe*, then we call the output constraints $\alpha$MOTe-Greedy* will be called G1 constraints or simply G1. 
\par 
Following \cite{phung-hamel-exact}, let $I_M(x,y)$ be the $M$-interference set consisting of all candidates that can appear between $y$ and $x$ in that order among all possible rankings that satisfy $M$. Fix an interference threshold $r\geq 0$ and suppose that $(x,y)$ is a pair with $|I_M(x,y)|\leq r$. Suppose that $y$ is not the winner in any median of the problem restricted to every union of $\{x,y\}$ with a subset of $ I_M(x,y)$. Then  $y$ cannot be ranked before $x$ in any median. In other words, we obtain $(x,y)$ as a valid constraints of the constrained problem. The resulting algorithm is  called LowInf-Pairs. Since the threshold $r$ is fixed, the runtime complexity of LowInf-Pairs is at most quadratic $O(n^2)$ as we may need to check all the pairs of candidates. In practice, we invoke  LowInf-Pairs after computing $\alpha$MOTe* or $\alpha$MOTe. 
   
\section{The main algorithm $\Gamma$-\textbf{\textup{XCC}}}
 \label{s:constant-approx-factor-gamma-xcc} 
 
In this section, we introduce the general algorithm $\bm \Gamma$-\textbf{XCC} 
  and study its theoretical properties. For any given set of  valid pairwise constraints $M \subset C^2$, the algorithm $\bm \Gamma$-\textbf{XCC} generalizes the well-known  Copeland algorithm  by devising a rigorous way to   incorporate $M$ into the computation of the score of each candidates.  
  \par 
Specifically, for every input weighted tournament over $C$ and every input set of constraints $M\subset C^2$,  the algorithm $\bm \Gamma$-\textbf{XCC}  produces a ranking $\pi^M$  which ranks candidates  $x\in C$ according to the following  decreasing indexes 
 (with ties broken arbitrarily)  
\begin{align}
\label{def:gamma-x-M-1}
\gamma_x^M & = |\{y \in C  \colon (x,y) \in M  \text{ {or else} } \delta_{xy}>0  \text{ and } (y,x)\notin M\}|\\
\label{def:gamma-x-M-2}
&=   \sum_{y \in C} \mathbf{1}((x,y)\in M  \vee (\delta_{xy}>0\wedge ((x,y), (y,x)\notin M))\\
\label{def:gamma-x-M-3}
& =   \sum_{y \in C\colon (x,y),(y,x)\notin M}  \langle a_{xy} \rangle +  \sum_{y \in C\colon (x,y)\in M} 1  
\end{align}
\noindent 
Here, $\langle b \rangle $ is  the round-down function defined by 
  $\langle b\rangle = 0$ if $0\leq b\leq \frac{1}{2}$ and  $\langle b  \rangle = 1$ if $\frac{1}{2}<b\leq 1$. 
We prove the following general approximation result for the generalization $\pmb \Gamma$-\textbf{\textup{XCC}}  of the Copeland  algorithm.

\begin{theorem}
\label{t:9-appr-gamma-xcc}
Let $M\subset C^2$ be a set of pairwise constraints. Suppose that $\pi^*$ is a median of the constrained Kemeny problem with input $M\subset C^2$. Let $\pi^M$ be the ranking output of  Algorithm $\bm \Gamma$-$\textbf{\textup{XCC}}$ with input $M$. 
 \begin{itemize} 
\item  For constrained weighted tournaments, Algorithm $\bm \Gamma$-$\textbf{\textup{XCC}}$ produces a 9-approximation of the Kemeny problem, i.e., $d_{KT}(\pi^M) \leq 9 d_{KT}(\pi^*)$. 
\item 
For constrained elections, Algorithm $\bm \Gamma$-$\textbf{\textup{XCC}}$ with input $M$ produces a 4-approximation of the Kemeny problem, i.e.,  $d_{KT}(\pi^M) \leq 4 d_{KT}(\pi^*)$. 
\end{itemize} 
\end{theorem}

\begin{algorithm}[hbt!]
\caption{$\bm \Gamma$\textbf{-XCC}}\label{algo:local-optimizer}
\DontPrintSemicolon 
\KwData{  $C$,  $[\delta_{xy}]_{x,y\in C}$, $M\subset C^2$} 
\KwResult{a ranking $\pi^M$}
$n \gets |C|$  \\ 
\For{$x\in C$}{$\gamma^M_x \gets \displaystyle  \sum_{y \in C} \mathbf{1}((x,y)\in M  \vee (\delta_{xy}>0\wedge ((x,y), (y,x)\notin M))$\; 
}  
$(\gamma_{x_1},\dots, \gamma_{x_n}) \gets \mathrm{Sort}\,  \{x\colon x\in C\}$ by decreasing values $\gamma_x$ \\ 
$\pi^M \gets x_1>x_2>\dots>x_n$ \\ 
\KwRet{$\pi^M$}
\end{algorithm} 
 
 As for  the Copeland  algorithm, the  time complexity of $\bm \Gamma$-\textbf{\textup{XCC}} is also $O(n^2)$ where $n=|C|$ since computing  $\gamma_x$ takes $O(n)$ for each $x \in C$ and sorting $n$ numbers takes $O(n\log n)$.  The proof of Theorem~\ref{t:9-appr-gamma-xcc} will be presented in the next two sections \ref{s:tournament-properties-Gamma-XCC}, \ref{s:election-properties-Gamma-XCC}. We then show in Section~\ref{s:axiomatic-properties-Gamma-XCC} that  $\bm \Gamma$-\textbf{\textup{XCC}} satisfies  several desired axiomatic properties such as the Condorcet winner criterion (Theorem~\ref{t:gamma-xcc-condorcet}), the Condorcet loser criterion (Theorem~\ref{t:gamma-xcc-condorcet-loser}), and more generally the Condorcet consistency property.    
\par 
For the proof  of Theorem~\ref{t:9-appr-gamma-xcc}, let us fix  a weighted tournament on a set of candidates $C$ with the weight matrix  $[a_{xy}]_{x,y \in C}$. 
For each $x \in C$, we define 
\begin{align}
\label{def:i-x-M-1} 
i_x & = \sum_{y\in C} a_{xy}\\ 
I_x & = (n-1) - i_x    
\end{align}
 \noindent 
A key ingredient for the proof of our constant-approximation factors is the following standard inequalities comparing the Kendall-tau distance and the well-known Spearman's footrule distance. For all rankings $\pi, \sigma$ over the same set of candidates $C$, the Spearman's footrule distance is given by 
\begin{align*}
 d_{S}(\pi, \sigma) = \sum_{x\in C} |\pi(x) - \sigma(x)|. 
\end{align*}

\begin{lemma}
\label{l:kemeny-spearman} 
$d_{KT}(\pi, \sigma) \leq d_S(\pi, \sigma) \leq 2d_{KT}(\pi, \sigma)$ for all rankings $\pi, \sigma$ over the same set of candidates. 
\end{lemma}

As a consequence, we have the following general comparison between the Spearman's footrule distance and the Kendall-tau distance.  

\begin{lemma}
\label{l:kendall-tau-vs-footnote}
 $d_S(\pi, \sigma)  \geq  d_{KT}(\pi, \sigma)  \geq  d_{KT}(\pi) - d_{KT}(\sigma)$. 
\end{lemma}

\begin{proof}
    See \cite[Lemma~3]{winners-ordering}. For ease of reading, we provide a proof.  Let $D$ be the set of  $\{x,y\}\in C^2$ such that $\mathrm{top}(\pi,{\{x,y\}})\neq \mathrm{top}(\sigma,{\{x,y\}})$. By Lemma~\ref{l:kemeny-spearman} and the triangle inequality, we have \begin{align*}
       d_S(\pi, \sigma)\geq  d_{KT}(\pi, \sigma)   =  |D| = \sum_{\{x,y\} \in D} 1 &\geq  \sum_{\{x,y\} \in D} |a_{xy}-a_{yx}| \\
        & \geq  d_{KT}(\pi) -  d_{KT}(\sigma). 
    \end{align*}
    The last inequality results from the fact that every unordered pairs $\{x,y\}\subset C$ contributes non trivially to $ d_{KT}(\pi) -   d_{KT}(\sigma)$ only if $\{x,y\}\in D$ and the contribution is $\pm (a_{xy}-a_{yx})$ which is bounded by $|a_{xy}-a_{yx}|$. 
\end{proof}

\begin{lemma}
\label{l:ordering-J-pi}
Let $H\colon C \to \mathbb{R}$ be a function. Let  $\pi^H$ be a ranking obtained by ordering increasingly the candidates $x\in C$ according to $H(x)$. Then for every ranking $\pi$ of $C$, we have  
    $\displaystyle \sum_{x\in C}|\pi(x) - H(x)| \geq \sum_{x\in C}|\pi^H (x) - H(x)|$. 
\end{lemma}

\begin{proof}
    This is a general properties saying that among all rankings $\rho$ of $C$, $\sum_{x\in C} |\rho (x) - f(x)|$ is minimized when $\rho = \pi^H$. The reader can consult \cite[Lemma~2]{winners-ordering} for the details. 
\end{proof}

\subsection{$\pmb \Gamma$-\textbf{\textup{XCC}} for constrained weighted tournaments}
\label{s:tournament-properties-Gamma-XCC}
In this section, we will prove the first part of Theorem~\ref{t:9-appr-gamma-xcc}. 
\begin{theorem}
\label{t:9-appr-gamma-alpha-xcc}
   The algorithm  $\bm \Gamma$-\textbf{\textup{XCC}} produces a 9-approximation of the Kemeny problem for constrained weighted tournaments. 
\end{theorem}

For the proof of the above theorem, let $M\subset C^2$ be the set of input constraints. Let $\pi^M$ be the ranking returned by $\bm \Gamma$-\textbf{\textup{XCC}} and let $\pi^*$ be a median of the associated constrained weighted election with constraints $M$. Let $n=|C|$. For every $x \in C$, we define    
\begin{equation}  J^M_x = (n-1)-\gamma^M_x,
\end{equation}
then $\pi^M$ is a ranking of $C$ according to the increasing order of $J_x^M$. 
We begin with the following auxiliary result.

\begin{lemma}
\label{l:constant-approx-lemma-2-alpha}
$\displaystyle \sum_{x \in C} |\pi^*(x) - J^M_x| \leq  4 d_{KT}(\pi^*)$.
\end{lemma}

\begin{proof}
Inspired by the arguments in \cite{winners-ordering}, let us define the following non-negative quantities 
\begin{align*}
    L^-_x & = \sum_{y \in C} \mathbf{1}( \pi^*(y)<\pi^*(x)) a_{yx} \\
    R^-_x & = \sum_{y \in C}  \mathbf{1}( \pi^*(y)> \pi^*(x)) a_{yx} \\
    L^+_x & = \sum_{y \in C}  \mathbf{1}( \pi^*(y)<\pi^*(x))  a_{xy}. 
\end{align*}

\noindent Since $a_{xy}+a_{yx}=1$ for all distinct $x,y \in C$,  we find that 
\begin{align}
\label{eq:proof-9-appro-2a}
    \pi^*(x) & =  L^-_x + L^+_x\\
\label{eq:proof-9-appro-2b}
    I_x & = L^-_x + R^-_x \\ 
    \label{eq:proof-9-appro-2c}
2 d_{KT}(\pi^*) & =  \sum_{x \in C} \left( L^+_x + R^-_x  \right). 
\end{align}
\noindent 
It follows from the triangle inequality that 
\begin{align}
\label{eq:proof-gamma-alpha-xcc-bound-ix}
    \sum_{x \in C}  |\pi^*(x)-I_x|  =   \sum_{x \in C} |L_x^+ -R_x^-| \leq     \sum_{x \in C} (L_x^+ +R_x^-)  =  2  d_{KT}(\pi^*). 
\end{align}
 
Let  $D^M_x = I_x - J^M_x=\gamma_x^M - i_x$. By the relations  $ \langle a_{xy}\rangle - a_{xy} = \pm \min (a_{xy}, a_{yx})$ and $a_{xy}+a_{yx}=1$ for all $x\neq y$, it follows from \eqref{def:gamma-x-M-3} and \eqref{def:i-x-M-1} that 
    \begin{align}
    \label{eq:proof-9-appro-2d-alpha}
        D^M_x  
        & =  \gamma_x^M -i_x   \nonumber\\
        & =   \sum_{y \in C\colon (x,y),(y,x)\notin M}  \langle a_{xy} \rangle +  \sum_{y \in C\colon (x,y)\in M} 1  \nonumber   \\
        &   -    \sum_{y \in C\colon (x,y),(y,x)\notin M}   a_{xy}  -  \sum_{y \in C\colon (x,y)\in M} a_{xy}  -    \sum_{y \in C\colon (y,x)\in M} a_{xy}    \nonumber \\      
& =
       \sum_{y \in C\colon (x,y),(y,x)\notin M} (\langle a_{xy}\rangle - a_{xy})\nonumber \\
        & +  \sum_{y \in C\colon (x,y) \in M} (1-a_{xy})   -  \sum_{y \in C\colon (y,x) \in M} a_{xy}  \nonumber \\
         & =  \sum_{y \in C\colon (x,y),(y,x)\notin M}  \pm \min (a_{xy}, a_{yx})  \nonumber \\
        &  +  \sum_{y \in C\colon (x,y) \in M}  a_{yx} -  \sum_{y \in C\colon (y,x) \in M} a_{xy}. 
    \end{align}
Summing $| D^M_x|$ over all $x\in C$, we obtain from \eqref{eq:proof-9-appro-2d-alpha} and the triangle inequality  the following estimation
\begin{align}
    \label{eq:proof-9-appro-2d-alpha-1}
    \sum_{x\in C}|  D^M_x| & \leq  \sum_{x,y\in C \colon (x,y),(y,x)\notin M} \min (a_{xy}, a_{yx})  \nonumber \\
        &  +  \sum_{x,y \in C\colon (x,y) \in M}  a_{yx} +   \sum_{x,y \in C\colon (y,x) \in M} a_{xy}  \nonumber \\
        & \leq 2 \sum_{\{x,y\}\subset C\colon  (x,y),(y,x)\notin M} ( \mathbf{1}(x>^{\pi^*} y) a_{yx} + \mathbf{1}(y>^{\pi^*} x) a_{xy}) \nonumber  \\
        & +   \sum_{\{x,y\} \in C\colon (x,y)\in M \vee (y,x) \in M} ( \mathbf{1}(x>^{\pi^*} y) a_{yx} + \mathbf{1}(y>^{\pi^*} x) a_{xy})   \nonumber          \\
       &   \leq 2 \sum_{\{x,y\}\subset C} ( \mathbf{1}(x>^{\pi^*} y) a_{yx} + \mathbf{1}(y>^{\pi^*} x) a_{xy}) \nonumber  \\
      &   =  2 d_{KT}(\pi^*). 
\end{align}
The second inequality follows from the trivial inequality  $\min (a_{xy}, a_{yx}) \leq  \mathbf{1}(x>^{\pi^*} y) a_{yx} + \mathbf{1}(y>^{\pi^*} x) a_{xy}$ and the fact that $  \mathbf{1}((x,y) \in M) a_{yx} +    \mathbf{1}((y,x) \in M)  a_{xy}  =  \mathbf{1}(x>^{\pi^*} y) a_{yx} + \mathbf{1}(y>^{\pi^*} x) a_{xy}$  if $(x,y)\in M \vee (y,x) \in M$, since  $M$ is a set of constraints satisfied by $\pi^*$. 
\par 
\noindent 
From the triangle inequality and the estimations \eqref{eq:proof-gamma-alpha-xcc-bound-ix}, \eqref{eq:proof-9-appro-2d-alpha-1},  we have  
\begin{align*}
    \sum_{x \in C} |\pi^*(x) - J^M_x| & = \sum_{x \in C} |\pi^*(x) - I_x + D^M_x| 
    \\
    & \leq \sum_{x \in C} |\pi^*(x) - I_x| + \sum_{x \in C}  |D^M_x|\nonumber\\
    & \leq 
2 d_{KT}(\pi^*) +  2 d_{KT}(\pi^*)\nonumber\\
    & =  4 d_{KT}(\pi^*)\nonumber
\end{align*}
whence the conclusion. 
\end{proof}

To complete the proof of Theorem~\ref{t:9-appr-gamma-alpha-xcc}, it suffices to apply  Lemma~\ref{l:constant-approx-lemma-2-alpha}, Lemma~\ref{l:ordering-J-pi}, the triangle inequality, and Lemma~\ref{l:kendall-tau-vs-footnote} in that order: 

\begin{align*}
8  d_{KT}(\pi^*) & \geq \sum_{x \in C} |\pi^*(x) - J^M_x| + \sum_{x \in C} |\pi^*(x) - J^M_x| \\
    & \geq \sum_{x \in C} |\pi^*(x) - J^M_x| + \sum_{x \in C} |\pi^M(x) - J^M_x|\\
    & \geq \sum_{x \in C} |\pi^*(x) - \pi^M(x)|  = d_S(\pi^*, \pi^M)\\
    & \geq  d_{KT}(\pi^M) -  d_{KT}(\pi^*).  
\end{align*}
Therefore, $d_{KT}(\pi^M) \leq 9 d_{KT}(\pi^*) $ and the proof of Theorem~\ref{t:9-appr-gamma-alpha-xcc}  is complete. 

\subsection{$\pmb \Gamma$-\textbf{\textup{XCC}} for constrained elections}
\label{s:election-properties-Gamma-XCC}
When restricted to the Kemeny problem for elections, the constant bound approximation of  $\pmb \Gamma$-\textbf{\textup{XCC}} can be significantly improved to 4 as follows. 

\begin{theorem}
\label{t:4-appr-gamma-alpha-xcc}
   The algorithm  $\bm \Gamma$-\textbf{\textup{XCC}} is a 4-approximation of  the Kemeny problem for constrained elections. 
\end{theorem}

\begin{proof}
Let $\sigma_1, \dots, \sigma_m$ be the collection of all votes in an election $(C,V)$ (repetition allowed) where $C$ is the set of candidates and $V$ is the voting profile. Let $\pi^M$ be the solution returned by Algorithm~ $\bm \Gamma$-\textbf{\textup{XCC}} with input constraints $M$. Let $\pi^*$ be a median of the constrained election such that $\pi^*$ satisfies all the pairwise constraints in $M$. Let  us denote the vector  $u=[u_{x,y}]_{x,y \in C}$ where  $u(x,y)= \mathbf{1}((x,y)\in M  \vee (\delta_{xy}>0\wedge ((x,y), (y,x)\notin M))$. In particular, observe that  $\gamma_x^M = \sum_{y \in C} u(x,y)$ and thus $J_x^M = (n-1)- \sum_{y \in C} u(x,y)$. Similarly, let   $u_i=[u_i(x,y)]_{x,y \in C}$ where  $u_i(x,y)= \mathbf{1}(\sigma_i(x) < \sigma_i(y))$. In particular, $u_i(x,x)=0$. We define also $v=[v(x,y)]_{x,y \in C}$ where $v(x,y)=  \mathbf{1}(\pi^*(x) < \pi^*(y))$.  Hence, if $(x,y) \in M$ then we have 
\begin{align}
\label{eq:election-gamma-xcc-0}
v(x,y)=u(x,y), \quad  v(y,x)=u(y,x)=0.
\end{align}  
\par 
\noindent 
 Recall that $\sigma_i(x)\in \{0,1,\dots, |C|-1\}$ denotes the ranking position of $x$ in $\sigma$. With the above notations, we have 
\begin{align*}
\sigma_i(x) =  \sum_{y \in C \setminus \{x\}} u_i (y,x) =  \sum_{y \in C \setminus \{x\}} (1-u_i (x,y))  = (n-1)-  \sum_{y \in C} u_i (x,y) , 
\end{align*}
and observe also that 
\begin{align}
\label{eq:election-gamma-xcc-1}
J_x^M - \sigma_i(x)  & =    (n-1)- \sum_{y \in C} u(x,y)  - (n-1) + \sum_{y \in C} u_i(x,y)  \nonumber \\
& =  \sum_{y \in C} (u_i(x,y) - u(x,y)). 
\end{align}

Let $M^{\perp} = \{(x,y) \in C^2\colon (x,y),(y,x)\notin M\}$ be the set of ordered pairs $(x,y)$ such that  neither $(x,y)$ nor $(y,x)$ is in $M$. Since each median is a minimizer of the sum of distance and $u(x,y)$ is a median of the values $u_1(x,y), \dots, u_m(x,y)$ whenever $(x,y) \in M^\perp$, we deduce that  
\begin{align} 
\label{eq:election-gamma-xcc-2}
 \sum_{i=1}^m \sum_{(x,y) \in M^{\perp}}  | u(x,y)  -  u_i(x,y)| & =    \sum_{(x,y) \in M^{\perp}}  \sum_{i=1}^m | u(x,y)  -  u_i(x,y)| \nonumber \\
 & \leq  \sum_{(x,y) \in M^{\perp}}  \sum_{i=1}^m | v(x,y)  -  u_i(x,y)| \nonumber \\ 
 &  =   \sum_{i=1}^m \sum_{(x,y) \in M^{\perp}}  | v(x,y)  -  u_i(x,y)|. 
\end{align}
\noindent 
We can now estimate  $d_{KT}(\pi^M) $ as follows. 

\begin{align*}
m d_{KT}(\pi^M) & = \sum_{i=1}^m d_{KT}(\pi^M, \sigma_i)  \leq \sum_{i=1}^m \sum_{x\in C}  |\pi^M(x) - \sigma_i(x) | & (\text{Lemma }\ref{l:kemeny-spearman})\\
& \leq \sum_{i=1}^m \sum_{x\in C}  \left(  |\pi^M(x) - J_x^M | +  | J_x^M - \sigma_i(x) |  \right) & (\text{triangle ineq.}) \\
& \leq 2  \sum_{i=1}^m \sum_{x\in C}  | J_x^M - \sigma_i(x) | & (\text{Lemma }\ref{l:ordering-J-pi})\\
& = 2  \sum_{i=1}^m \sum_{x\in C}  |\sum_{y \in C} (u(x,y)  - u_i(x,y))|  & (\eqref{eq:election-gamma-xcc-1})\\
& \leq  2  \sum_{i=1}^m \sum_{(x,y) \in C^2}  | u(x,y)  -  u_i(x,y)| & (\text{triangle ineq.}) 
\end{align*}
The last sum can be expressed and bounded as follows
\begin{align*}
&   2  \sum_{i=1}^m \sum_{(x,y) \in M^{\perp}}  | u(x,y)  -  u_i(x,y)|    +  2  \sum_{i=1}^m \sum_{(x,y) \in  M}  | u(x,y)  -  u_i(x,y)|  \\
 & \leq   2  \sum_{i=1}^m \sum_{(x,y) \in M^{\perp}}  | v(x,y)  -  u_i(x,y)|  & (\eqref{eq:election-gamma-xcc-0}) \\
 & +  2  \sum_{i=1}^m \sum_{(x,y) \in  M}  | v(x,y)  -  u_i(x,y)|   &  (\eqref{eq:election-gamma-xcc-2})     \\
 & =  2  \sum_{i=1}^m \sum_{(x,y) \in C^2}  | v(x,y)  -  u_i(x,y)|= 4md_{KT}(\pi^*). 
 \end{align*}
We conclude that $d_{KT}(\pi^M) \leq 4 d_{KT}(\pi^*)$ and the conclusion follows. 
\end{proof}

\subsection{Some axiomatic properties of $\pmb  \Gamma$-$\textbf{\textup{XCC}}$}
\label{s:axiomatic-properties-Gamma-XCC}
We now turn our  attention to some important axiomatic properties of $\bm \Gamma$-\textbf{\textup{XCC}}.  
In a given weighted tournament, a candidate $x$ is a \textbf{Condorcet winner} if for every other candidate $y$ a strict majority  of votes rank $x$ before $y$, i.e., $\delta_{xy}=a_{xy} - a_{yx} >0$. It is clear that a Condorcet winner when exists must be unique.  
We say that an aggregation algorithm satisfies the \textbf{Condorcet winner criterion} if every  Condorcet winner, when exists, must be top-ranked by the algorithm.  An aggregation algorithm satisfies the \textbf{XCC consistency property} if for every weighted tournament over $C$, it returns a ranking $\pi$ which respects the finest XCC partition $X_1\coprod X_2 \dots \coprod X_k$ of the tournament, i.e., $x_i>^\pi x_j$ for all $x_i\in X_i$ and $x_j \in X_j$ with $i<j$.  
Note that the XCC consistency property implies the Condorcet winner criterion. Moreover, 
every algorithm computing an exact median of a weighted tournament satisfies the XCC consistency property. 
  \par 
The strongest property that we consider is the following. We say that an aggregation algorithm satisfies the \textbf{M-XCC consistency property} if for every constrained weighted tournament over $C$ with input constraints $M\subset C^2$, it returns a ranking $\pi$ which respects the finest $M$-XCC partition $X_1\coprod X_2 \dots \coprod X_k$ of that constrained weighted tournament, i.e., $x_i>^\pi x_j$ for all $x_i\in X_i$ and $x_j \in X_j$ with $i<j$. 
 \par 
The following theorem shows that $\bm \Gamma$-$\textbf{\textup{XCC}}$ is placed among a few significant approximation algorithms which satisfy the Condorcet winner criterion such as Soft Condorcet Ranking \cite{soft-condorcet}.  

\begin{theorem}
\label{t:gamma-xcc-condorcet}
    Suppose that $x$ is a Condorcet winner  of a weighted tournament over a set of candidates $C$. Let $M\subset C^2$ be a set of constraints satisfied by some median $\pi^*$ of the weighted tournament.  
 Then $x$ is the winner in every  ranking $\pi^M$ returned by $\bm \Gamma$-$\textbf{\textup{XCC}}$ with input $M$.  
 \end{theorem}

\begin{proof}
 
Since $\pi^*$ is a median, $x$ must be the winner of the ranking $\pi^*$. It follows that $(z,x)\notin M$ for all  $z \in C \setminus \{x\} $ since $\pi^*$ satisfies all the constraints in $M$ by hypothesis. 
 Moreover, $\delta_{xz}>0$ for all $z \in C \setminus \{x\}$  because $x$ is a Condorcet winner.  We deduce from \eqref{def:gamma-x-M-1} that  
\begin{align*}
\gamma_x^M   & = |\{z \in C  \setminus \{x\}\colon (x,z) \in M  \text{ {or else} } \delta_{xz}>0  \text{ and } (z,x)\notin M\}| \\
& =  |\{z \in C \setminus \{x\}  \colon (x,z) \in M  \text{ {or else} } \delta_{xz}>0 \}| \\
& = |\{z \colon z \in C \setminus \{x\}\}| \\
& = |C|-1. 
\end{align*}

\noindent 
Let $y \in C \setminus \{x\}$. We have shown above that $(y,x)\notin M$ and $\delta_{yx}<0$. Hence, $x \notin \{z \in C  \setminus \{y\}\colon (y,x) \in M  \text{ {or else} } \delta_{yx}>0  \text{ and } (x,y)\notin M\}$ and by   \eqref{def:gamma-x-M-1},  
\begin{align*}
\gamma_y^M   & = |\{z \in C  \setminus \{x\}\colon (x,z) \in M  \text{ {or else} } \delta_{xz}>0  \text{ and } (z,x)\notin M\}| \\
& \leq |C\setminus \{x,y\}|  =|C|-2   \\
& <  \gamma_x^M =  |C|-1
\end{align*} 
 
Having the largest index $\gamma_x^M$ without ties to any other candidate, $x$ must be the winner in every ranking $\pi^M$ returned by $\bm \Gamma$-$\textbf{\textup{XCC}}$. 
\end{proof}

Moreover,  we show that $\bm \Gamma$-$\textbf{\textup{XCC}}$ also satisfies the Condorcet loser criterion. A candidate $x$ is called a Condorcet loser in a weighted tournament  with weight matrix $[a_{xy}]_{x,y \in C}$ over a set of candidates $C$ if $\delta_{xy} = a_{xy} - a_{yx}< 0$ for all $y \in C\setminus \{x\}$. An aggregation algorithm satisfies the \textbf{Condorcet loser  criterion} if every  Condorcet loser, when exists, must be bottom-ranked by the algorithm.   

\begin{theorem}
\label{t:gamma-xcc-condorcet-loser}
  Suppose that $x$ is a Condorcet loser  in a weighted tournament over a set of candidates $C$. Let $M\subset C^2$ be a set of constraints satisfied by some median $\pi^*$ of the weighted tournament.  
 Then $x$ is the loser in every  ranking $\pi^M$ returned by $\bm \Gamma$-$\textbf{\textup{XCC}}$ with input $M$.  
 \end{theorem}

\begin{proof}
Since $\pi^*$ is a median of the weighted tournament, $x$ must be the loser in the ranking $\pi^*$. It follows that $(x,z)\notin M$ for every  $z \in C \setminus \{x\} $ since $\pi^*$ satisfies all the constraints in $M$ by hypothesis. Moreover, $\delta_{xz}=a_{xz}-a_{zx} <0$ for all $z \in C \setminus \{x\}$  by hypothesis. 
Consequently, we infer from  \eqref{def:gamma-x-M-1} that 
\begin{align*}
\gamma_x^M   = |\{z \in C  \colon (x,z) \in M  \text{ {or else} } \delta_{xz}>0  \text{ and } (z,x)\notin M\}| = 0. 
\end{align*}
\noindent 
 Let $y \in C \setminus \{x\}$. We have $  \delta_{yx}  > 0$ and  either $(y,x)\in M$ or $(y,x) \notin M$ (note that $(x,y)\notin M$ as shown above). Consequently, $x\in \{z \in C  \colon (y,z) \in M  \text{ {or else} } \delta_{yz}>0  \text{ and } (z,y)\notin M\}$ and thus by  \eqref{def:gamma-x-M-1}, we have 
\begin{align*}
 \gamma_x^M & = |\{y \in C  \colon (x,y) \in M  \text{ {or else} } \delta_{xy}>0  \text{ and } (y,x)\notin M\}| \geq 1 >  0 = \gamma_x^M.
\end{align*}   
Having the lowest index $\gamma_x^M=0$ without ties to any other candidate, $x$ must be the loser in every ranking $\pi^M$ returned by $\bm \Gamma$-$\textbf{\textup{XCC}}$.  
\end{proof}

More generally, we prove the following M-XCC consistency properties for our Algorithm  $\bm \Gamma$-$\textbf{\textup{XCC}}$. 
    
\begin{theorem}
\label{t:xcc-consistency}
$\bm \Gamma$-$\textbf{\textup{XCC}}$ satisfies the  M-XCC consistency property. More specifically, let  $M\subset C^2$ be a set of constraints satisfied by some median $\pi^*$ of a given constrained weighted tournament.  Then the algorithm $\bm \Gamma$-$\textbf{\textup{XCC}}$ with input $M$ returns a ranking $\pi^M$ which satisfies the finest M-XCC-partition. 
\end{theorem}

\begin{proof}
This is a direct application of \cite{phung-hamel-exact}[Algorithm 6] which computes the finest $M$-XCC partition (see Section~\ref{s:M-ACP} below for some details).  \cite{phung-hamel-exact}[Theorem 9] on the correctness of the mentioned algorithm computing the finest $M$-XCC partition implies that all elements $x$ with the same indices $\gamma^M_x$ must belong to the same constituting block of the finest $M$-XCC-partition. Since $\bm \Gamma$-$\textbf{\textup{XCC}}$ with input $M$ orders the elements $x$ of $C$ by their decreasing indices $\gamma^M$, it follows that $\pi^M$ satisfies the finest $M$-XCC-partition.
\end{proof}

\section{Constrained Approximate Condorcet Partitioning M-ACP}
 \label{s:M-ACP}
In this section, we describe and analyze our generalization \textbf{M-ACP} of the \textbf{ACP} method of \cite{escobedo}. 
We first recall the notion of M-Condorcet partitions introduced in \cite{phung-hamel-exact}, adapted in our context of weighted tournament.

\begin{definition}  
\label{def:M-XCC-partition} 
 Consider a constrained weighted tournament over $C$ with weighted matrix $[a_{xy}]_{x,y \in C}$ and let $\delta_{xy}=a_{xy}-a_{yx}$. Let $M \subset C^2$.   
 We say that a partition $C=\bigcup_{i=1}^k X_i$, also denoted $\{X_1,\cdots, X_k\}$,    is an $M$-Condorcet  partition ($M$-XCC partition) if $(x_i,x_j)\in M$ or else $\delta_{x_ix_j}>0$ and $(x_j,x_i)\notin M$  for all $x_i\in X_i$ and $x_j \in X_j$ with $i<j$. Such a partition with most blocks is called a finest $M$-Condorcet partition.  
\end{definition}

By results in \cite{phung-hamel-exact}, if $\{X_1,\cdots, X_k\}$ is an $M_\pi$-XCC partition where the set of constraints $M_\pi$ is satisfied by some  median $\pi$ of the constrained weighted tournament, then $\pi$ must respect this partition, that is, $x_i>^\pi x_j$ for all $x_i\in X_i$, $x_j \in X_j$ with $i<j$. Moreover, the authors  prove that if $M\subseteq C^2$ is an anti-symmetric set of constraints, then the finest $M$-Condorcet partition exists and is unique.  They also provided a quadratic time complexity $O(|C|^2)$ algorithm that outputs the finest $M$-XCC partition.

 \subsection{Approximate  \textbf{M-ACP} partitions and solutions}

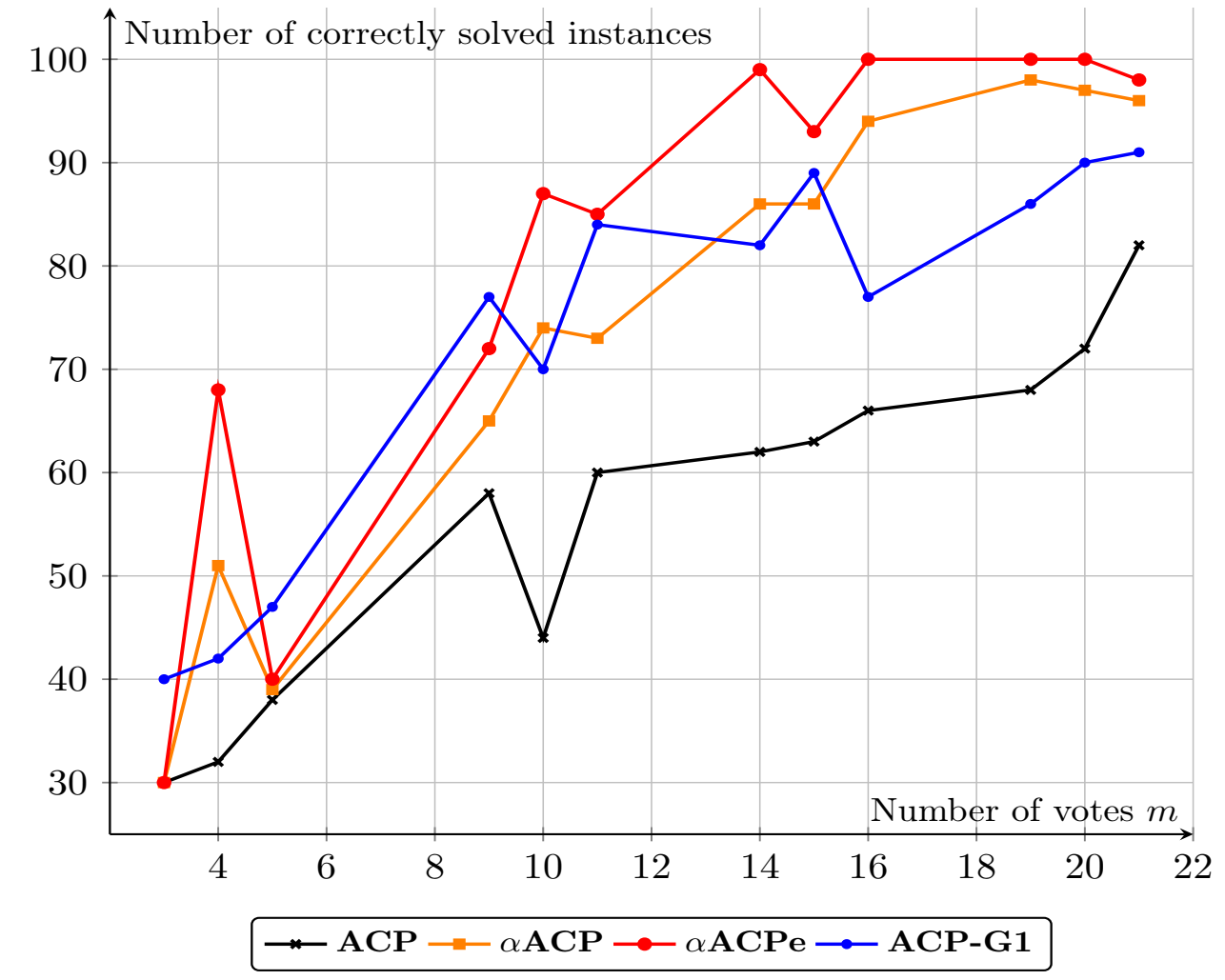
\begin{figure}[H] 
    \centering 
    \resizebox{8.6 cm}{6.7cm}{ 
    \begin{tikzpicture} 
        \begin{axis}[
            xlabel={\small Number of votes $m$},
            ylabel={\small  Number of correctly solved instances},
            legend style={font=\footnotesize, rounded corners=2pt, 
                at={(0.5,-0.1)}, 
                anchor=north,    
                legend columns=4 
            },
            grid=major, 
            enlarge x limits=false,
            ymax=105, 
            ymin=25, 
            xmax=22,
            xmin=2,  
            xlabel style={yshift=3pt}, 
            ylabel style={yshift=-5pt}, 
            axis lines=center,
        ]
            \addplot[color=black, line width =1pt, mark=x,mark size = 1.75pt] coordinates {(3,30) (4,32) (5,38) (9,58) (10,44) (11,60) (14,62) (15,63) (16,66) (19,68) (20,72) (21,82)};
            \addlegendentry{\textbf{ACP}}
            
            \addplot[color=orange,line width =1pt, mark=square*, mark size = 1.2pt] coordinates {(3,30) (4,51) (5,39) (9,65) (10,74) (11,73) (14,86)(15,86) (16,94)  (19,98) (20,97) (21,96)};
            \addlegendentry{$\alpha$\textbf{ACP}}

            \addplot[color=red,line width =1pt,mark=*,mark size = 1.5pt] coordinates {(3,30) (4,68) (5,40) (9,72) (10,87) (11,85) (14,99) (15,93) (16,100) (19,100) (20,100) (21,98)};
            \addlegendentry{$\alpha$\textbf{ACPe}}

            \addplot[color=blue,line width =1pt, mark=*,mark size = 1pt] coordinates {(3,40) (4,42) (5,47) (9,77) (10,70) (11,84) (14,82) (15,89) (16,77) (19,86)  (20,90) (21,91)};
            \addlegendentry{\textbf{ACP-G1}}

        \end{axis} 
    \end{tikzpicture}
    \vspace{-0.4cm}  
    }
    \caption{\small  Number of correctly solved instances by $M$-ACP methods on 100 random Mallows voting profiles with 50 candidates, $m\in \{3,4,5,9,10,11,14,15,16,19,20,21\}$ votes,   dispersion parameter $\theta=0.9$, interference threshold $r=10$, and approximation threshold $h=24$. }
   \label{fig:correct-instances} 
\end{figure}

In parallel to the approximation version  \textbf{ACP} with threshold $h$ of the Finest-Condorcet Partitioning algorithm described in \cite{escobedo}, we describe our constrained Approximate Condorcet Partitioning  \textbf{M-ACP}  (see Algorithm~\ref{alg:M-ACP}) of M-XCC partitioning.  {M-ACP} partitions contain in general more blocks than M-XCC partitions.  Moreover, experimental results show that we can consistently achieve high quality solutions from M-ACP partitions (see  Figure~\ref{fig:correct-instances} and Table~\ref{table:n=50,theta=0.9}).  
\par 
Let  $M\subseteq C^2$ be a set of constraints and let  $h>0$ be a  parameter which bounds the size of the blocks. We define the approximate $M$-ACP partition with threshold $h$ as follows.    
 Essentially, Algorithm~\ref{alg:M-ACP} takes as input the finest $M$-Condorcet partition $\textbf{X}^{f,M}$ and its initial partition $\textbf{X}^{0,M}$,  
decomposes blocks in $\textbf{X}^{f,M}$ if $|\textbf{X}^{f,M}|\geq h$ into sub-blocks from $\textbf{X}^{0,M}$, and  repeatedly merge as many as possible these consecutive sub-blocks without breaking the size limit $h$. As for   \textbf{ACP} and  \textbf{M-XCC}, the time complexity of  \textbf{M-ACP} is also quadratic in the number $n$ of candidates.

 \begin{algorithm}[ht!]
\caption{M-ACP}
\label{alg:M-ACP}
\DontPrintSemicolon 
\KwData{ 
 $C$, matrix $[\delta_{xy}]$, constraints $M\subseteq C^2$, threshold $h$} 
\KwResult{$M$-approximate Condorcet partition $\textbf{X}_h^{M\text{-}ACP}$} 
$\textbf{X}^{f,M}, \textbf{X}^{0,M} \gets \text{M-Finest-Condorcet Partition} ([\delta_{xy}], M)$\; 
$\textbf{X}_h^{M\text{-}ACP} \gets \varnothing$\;
\For {$k=1$ \textbf{ to } $|\textbf{X}^{f,M}|$}
{\If{$|\textbf{X}_k^{f,M}| \leq h$}
{Append $\textbf{X}_k^{f,M}$ to $\textbf{X}_h^{M\text{-}ACP}$ \; 
}
\Else{Let $X_u^0,\dots, X_v^0$ be the   consecutive subsets of $\textbf{X}^{0,M}$ that have been merged to form $\textbf{X}_k^{f,M}$ \;
$s \gets u$\; 
\While{$s \leq v-1$}
{\If {$|{X}_s^{0,M}| > h$ \textbf{or} $s=v-1$} 
{Append $X_s^{0,M}$ to $\textbf{X}_h^{M\text{-}ACP}$\; 
$s \gets s+1$\; }
\Else{For $t\leq v$ the highest index such that $|X_s^{0,M}\cup \cdots \cup X_t^{0,M} | \leq h$, 
merge $X_s^{0,M},  \dots , X_t^{0,M}$ and append it to $\textbf{X}_h^{M\text{-}ACP}$\; 
$s \gets t+1$}
}
}
}
\KwRet{$\textbf{X}_h^{M\text{-}ACP}$} 
\end{algorithm}
 
 For our experiments for  the \textbf{M-ACP} method over unconstrained elections (see  Figure~\ref{fig:correct-instances} and Table~\ref{table:n=50,theta=0.9}), let $\alpha$XCC, $\alpha$XCCe, and XCC-G1 denote
 the $M$-XCC partitions where $M$  is the set of constraints $\alpha$MOT,  $\alpha$MOTe, and  G1 respectively. Similarly, let $\alpha$ACP, $\alpha$ACPe, and ACP-G1 denote
 the $M$-ACP partitions where $M$  is the set of constraints $\alpha$MOT,  $\alpha$MOTe, and  G1 respectively. 
Let us also define 
{$M$-ACP solution with threshold $h$.} 
Once obtained  from Algorithm~\ref{alg:M-ACP} a partition    $\textbf{X}_h^{M\text{-}ACP} =\{ X_1,\dots, X_k\}$, we    
\begin{itemize}
\item  
find an exact solution of the subproblem $X_i$ if $|X_i| \leq h$; 
\item  permute randomly the candidates in $X_i$ if $|X_i| > h$;  \item  concatenate these sub-rankings into a full ranking $\pi$, which we call an $M${-ACP solution with threshold $h$}. 
\end{itemize}
By abuse of notation and when there is no risk of confusion, we also use $\alpha$ACP, $\alpha$ACPe, and ACP-G1 to denote the corresponding $M$-ACP solution with threshold $h$. 
We measure the standard {score~gap} to quantify the quality of the approximate solutions $M$-ACP (see Table~\ref{table:n=50,theta=0.9}): 
$$
\text{Score gap} = \frac{\text{Kemeny score of the approximate solution}}{\text{Kemeny score of medians}} - 1. 
$$
 The  {number of instances correctly solved} (with zero score gap) by $M$-ACP solutions is reported both in Figure~\ref{fig:correct-instances} and Table~\ref{table:n=50,theta=0.9}. Other performance indicators such as the density of pairwise constraints obtained ($\%$ pairs solved for $\alpha$MOT,  $\alpha$MOTe and $\%$ pairs for G1) out of the maximal $n(n-1)/2$ pairs, the number of instances where G1 constraints are all valid ($\#$ valid inst.), and the number of blocks ($\#$ blocks) and max block size in the various partitions are contained in   Table~\ref{table:n=50,theta=0.9}. 
  
\par 
\vspace{0.2cm} 

  \begin{table}[H]
    \centering
 \caption{\small Mean values for  $\%$ pairs solved by $\alpha$MOT and $\alpha$MOTe,   $\%$ pairs found by G1, $\#$ blocks and max block size of the partitions, score gap (gap), $\#$ valid instances ($\#$ valid inst.) for G1, and the number of correctly solved instances ($\#$ optimal) by M-ACP methods on 100 random Mallows voting profiles with $n\in \{30,40,50\}$ candidates, $m \in \{3,4,5,10,15,20\}$ votes, dispersion parameter $\theta =0.9$,  approximation threshold $h\in \{12,16,20\}$, and interference threshold $r=10$ for LowInf-Pairs  applied to  each method $\alpha$MOT, $\alpha$MOTe and G1. The best M-ACP solutions on average are highlighted for each triple $(n,m,h)$.}
 \label{table:n=50,theta=0.9}
   \resizebox{12.7 cm}{7.2cm}{  
\begin{tabular}{|c|c| ccccc|ccccc|}
      \hline \multicolumn{2}{|c|}{$n$} & \multicolumn{5}{c|}{$40$} & \multicolumn{5}{c|}{$50$}   \\
           \hline 
\multicolumn{2}{|c|}{$m$}      & 4&5 & 10& 15&20   & 4&5 & 10& 15&20\\
       \hline 
       \multicolumn{12}{ |c| }{\textbf{Qualities of constraints and partitions}}\\   
       \hline 
{$\text{\small $\%$ pairs}$}  \multirow{2}{*} & $\text{\small$\alpha$MOT}$        & 81.9& 75.7& 90.0& 92.7&94.2    & 81.8& 78.5& 91.3& 93.4 &95.6\\
 
{$\text{\small solved}$} &  $\text{\small$\alpha$MOTe}$   & 90.1& 80.1& 93.9& 95.0&97.0    & 89.1& 82.0 & 94.5& 95.8 &97.8\\
 \hline 
{$\text{\small $\%$ pairs}$} &  $\text{\small G1}$        &99.9 &85.6 &99.0 &96.6&99.3    & 99.8& 86.9& 99.1& 97.3&99.5\\
\hline 

 {\small \text{$\#$ valid inst.}} & $\text{\small G1}$    &65  & 100  & 79 & 96 & 88         & 42  & 100  & 70 & 91  & 90 \\
\hline  
 
 \multirow{4}{*}{$\text{\small $\#$  blocks}$} & $\text{\small XCC}$         & 1.31& 3.44& 1.79& 5.01&2.97    & 1.25& 3.43& 1.75& 5.51&3.13\\
 
&  $\text{\small$\alpha$XCC}$   & 4.23& 3.54& 8.05& 11.39&14.4    & 4.00& 3.63& 8.97& 12.08&17.4\\
 
& $\text{\small$\alpha$XCCe}$    & 11.61 & 4.55& 16.24&18.62&24.74    & 10.94& 4.75& 17.47& 17.71&28.53\\
 
&  $\text{\small XCC-G1}$       &39.82 &9.63 & 35.96& 26.22&36.65    & 48.49& 10.03& 37.07&23.61 &37.6\\
\hline 

\multirow{4}{*}{\small \text{block size}} & $\text{\small XCC}$    &39.65 & 36.59 & 38.93& 28.71&34.22    & 49.74 & 47.03 & 49.17& 35.02&40.89\\
 
{\small \text{max}}&  $\text{\small$\alpha$XCC}$       & 33.13& 36.34& 24.72&18.46 &15.05    & 43.33& 46.69& 30.40& 21.76&15.59\\
 
& $\text{\small$\alpha$XCCe}$      & 23.12& 34.99& 16.76&14.44&10.01    & 32.88& 44.95 &20.53& 16.87&10.44\\
 
  &  $\text{\small XCC-G1}$       & 1.18&29.74 & 4.78& 11.03&4.17    & 2.51& 37.16& 6.80& 12.51&5.57\\
\hline  
\multicolumn{12}{ |c| }{\textbf{Qualities of approximate solutions}}\\   
\hline 
 
\multirow{4}{*} & $\text{\small ACP}$    &6.254 & 2.445 & 1.502 & 0.568 &0.513     & 6.358  & 2.958 & 1.515 & 0.741 & 0.537\\
 
{\small \text{gap}  ({\tiny  \permille})}&  $\text{\small$\alpha$ACP}$       & 3.168 & 2.375 & 0.514& 0.343 &0.121            & 3.707& 2.930& 0.882 & 0.324 &0.069\\
 
\text{\small h=12} & $\text{\small$\alpha$ACPe}$      & 1.931 & 2.159 &  0.307 &2.203 &\cellcolor{orange}0.060        & \cellcolor{orange}2.089 & 2.784 &0.537 & 0.200& \cellcolor{orange}0.033\\
 
  &  $\text{\small ACP-G1}$      & \cellcolor{orange}1.293 & \cellcolor{orange}1.519  & \cellcolor{orange}0.303 & \cellcolor{orange}0.166 &0.088        & 2.175& \cellcolor{orange}2.372 & \cellcolor{orange}0.386& \cellcolor{orange}0.112 &0.040\\
\hline  

\multirow{4}{*} & $\text{\small ACP}$    &  10  & 17    &   31  &   42  &    50      & 0   & 7  & 14 &  18  & 36 \\
 
{\small $\#$ \text{optimal}}&  $\text{\small$\alpha$ACP}$       & 32   & 18   & 63  &  54  &         82  & 0  & 7 & 34 & 48 &    82 \\
 
{\small \text{h=12}}  & $\text{\small$\alpha$ACPe}$      &  53   &   22 & \cellcolor{orange}79  &   70   &  \cellcolor{orange}93   &   0 & 8   &   55  &  65  &   \cellcolor{orange}92 \\
 
  &  $\text{\small ACP-G1}$       & \cellcolor{orange}65   &   \cellcolor{orange}30   & 76    &  \cellcolor{orange}74  &   87  & \cellcolor{orange}8 & \cellcolor{orange}12 &  \cellcolor{orange}65   &  \cellcolor{orange}77  &  90 \\
\hline

\multirow{4}{*}  & $\text{\small ACP}$    & 4.556   & 1.693  & 1.230 & 1.499 & 0.260        & 4.579  & 1.578  & 1.089& 0.446 & 0.321 \\
 
{\small \text{gap} ({\tiny  \permille})}&  $\text{\small$\alpha$ACP}$       & 2.306 & 1.639 & 0.409 & 1.526  & 0.022        & 2.644 & 1.596 & 0.504& 0.135& 0.050\\
 
\text{\small h=16} & $\text{\small$\alpha$ACPe}$      & \cellcolor{orange}1.274 & 1.528 & \cellcolor{orange}0.141 & 1.522 & \cellcolor{orange}0     &\cellcolor{orange}1.904 & 1.505  & \cellcolor{orange}0.240 & 0.083 & \cellcolor{orange}0 \\
 
  &  $\text{\small ACP-G1}$       & 1.293 & \cellcolor{orange}1.414  & 0.246  & \cellcolor{orange}1.337 & 0.065   & 2.119 & \cellcolor{orange}1.302 & 0.284 & \cellcolor{orange}0.071&   0.040\\
\hline  
\multirow{4}{*}& $\text{\small ACP}$    &18   & 32  & 29  & 53  & 59        & 11  & 22  & 28 &  42  & 50 \\
 
{\small $\#$ \text{optimal}}&  $\text{\small$\alpha$ACP}$       & 50 & 34 & 74 & 81  & 94         & 33 & 23 & 54 & 70 &91\\
 
{\small \text{h=16}}  & $\text{\small$\alpha$ACPe}$      &  \cellcolor{orange}71 & 37 & \cellcolor{orange}90  & \cellcolor{orange}90  &  \cellcolor{orange}98   &  41 & \cellcolor{orange}26  & \cellcolor{orange}77  &  83  &  \cellcolor{orange}100 \\
 
 &  $\text{\small ACP-G1}$       & 65 &   \cellcolor{orange}39  & 78  &  89 & 88   & \cellcolor{orange}42 & \cellcolor{orange}26 &  69  &  \cellcolor{orange}83 &  90 \\
\hline

\multirow{4}{*} & $\text{\small ACP}$    &2.808  & 1.219  & 0.622 & 0.190 & 0.260        & 3.834  & 1.579  & 1.032& 0.342 &0.263 \\
 
{\small \text{gap ({\tiny  \permille})}}&  $\text{\small$\alpha$ACP}$       & 1.367 & 1.208 & 0.335 & 0.036  & 0.022        & 2.234 & 1.548 & 0.334& 0.089& 0.020\\
 
\text{\small h=20}  & $\text{\small$\alpha$ACPe}$      & \cellcolor{orange}1.245 & 1.241 & \cellcolor{orange}0.114 & 0.037 & \cellcolor{orange}0     &\cellcolor{orange}0.969 & 1.604  & \cellcolor{orange}0.183 & 0.058 & \cellcolor{orange}0.010 \\
 
 &  $\text{\small ACP-G1}$       & 1.293 & \cellcolor{orange}0.995  & 0.237  & \cellcolor{orange}0.025 & 0.065   & 2.101 & \cellcolor{orange}1.205 & 0.277 & \cellcolor{orange}0.054&   0.040\\
\hline  

\multirow{4}{*} & $\text{\small ACP}$    &49   & 40  & 56  & 70  & 72        & 19  & 19  & 24 &  48  & 66 \\
 
{\small $\#$ \text{optimal}}&  $\text{\small$\alpha$ACP}$       & 69 & 41 & 76 & 92  & 97         & 40 & 20 & 71 & 79 &96\\
 
\small \text{h=20} & $\text{\small$\alpha$ACPe}$      &  \cellcolor{orange}74 & 41 & \cellcolor{orange}91  & \cellcolor{orange}94  &  \cellcolor{orange}100   &  \cellcolor{orange}65 & 21  & \cellcolor{orange}82  &  85  &  \cellcolor{orange}98 \\
 
   &  $\text{\small ACP-G1}$       & 65 &   \cellcolor{orange}50  & 79  &  95 & 88   & 42 & \cellcolor{orange}35 &  70  &  \cellcolor{orange}88 &  90 \\
\hline  

\end{tabular}

   } 
  \end{table}

  \subsection{Provable  guarantees}

While $\alpha$MOT* based  \textbf{ACP} methods such as $\alpha$ACP, $\alpha$ACPe, and ACP-G1 {enjoy a constant-factor approximation for the Kemeny problem (see Section~\ref{s:constant-approx-factor-gamma-xcc}), the constant bound obtained is not satisfying for high stake applications. For this, we provide an easy to  compute instance-specific bounds on the distance between the obtained solution and any medians of the problem. Notably, this bound is achieved without computing the medians nor their Kemeny score. We rely on a set of constraints satisfied  by some median (such as  $\alpha$MOT* and $\alpha$MOTe* constraints) and the ability to obtain optimal solutions of instances of size at most $h$ (a parameter chosen manually). 

\begin{theorem}
\label{t:provable-guarantee}
Let $\pi$ be median of constrained weighted tournament with input constraints $W\subset C^2$. Let $M_\pi$ be a set of {constraints} satisfied by $\pi$ such that $W\subset M$.  Let $\{X_1,\cdots, X_k\}$ be an  $M_\pi$-Condorcet partition. Let $\hat{\pi}$ be any $M_\pi$-ACP solution with threshold $h$ which (i) solves to optimality each subset $X_i$ with $|X_i|\leq h$ and (ii) 
respects all the constraints imposed by $M_\pi$. 
Then 
\begin{align}
\label{e:guarantees}
    0 & \leq d_{KT}(\hat{\pi}, V) - d_{KT}( \pi,V)  \\&  \leq \sum_{p=1}^k \sum_{\,\{x,y \}\subset X_p} \mathbf{1}({(x,y), (y,x)\notin M_\pi})  (\delta^+_{xy} \mathbf{1}({y>^{\hat{\pi}}x}) +  \delta^+_{yx} \mathbf{1}({x>^{\hat{\pi}}y})) \nonumber 
\end{align} 
where $\delta_{xy}^+=\max(0,\delta_{xy})$ for all $x,y \in C$.  
\end{theorem}

\begin{proof}
    Since both $\pi$ and $\hat{\pi}$ respect the partition $\{X_1,\dots,  X_k\}$, no pairs $\{x,y\}$,  where $x\in X_i$ and $y\in X_j$ with $i\neq j$, can contribute non trivially to $d_{KT}(\hat{\pi}, V) - d_{KT}( \pi,V)$. If $x,y \in X_i$ with $|X_i|\leq h$ then the pair $\{x,y\}$ contribute nothing to $d_{KT}(\hat{\pi}, V) - d_{KT}( \pi,V)$ since both $\pi$ and $\hat{\pi}$ solve $X_i$ to optimality by hypothesis. Suppose now that $x,y\in X_i$ with $|X_i|> h$. If $x>^{\hat{\pi}}y$,  the pair $\{x,y\}$ contributes at 
    most $\delta^+_{yx}$ to $d_{KT}(\hat{\pi}, V) - d_{KT}( \pi,V)$. Similarly,   $\{x,y\}$ contributes at 
    most $\delta^+_{xy}$ if $y>^{\hat{\pi}}x$. Therefore,  the pair $\{x,y\}$ contributes at 
    most $(\delta^+_{xy} \mathbf{1}({y>^{\hat{\pi}}x}) +  \delta^+_{yx} \mathbf{1}({x>^{\hat{\pi}}y}))$ to $d_{KT}(\hat{\pi}, V) - d_{KT}( \pi,V)$. As  both $\hat{\pi}$ and $\pi$ respect the constraints $M_\pi$,  the pair $\{x,y\}$ contributes nothing when $(x,y)\in M_\pi$ or $(y,x)\in M_\pi$. Consequently, the pair $\{x,y\}$ can contribute at most    $\mathbf{1}({(x,y), (y,x)\notin M_\pi})  (\delta^+_{xy} \mathbf{1}({y>^{\hat{\pi}}x}) +  \delta^+_{yx} \mathbf{1}({x>^{\hat{\pi}}y}))$ to the difference $d_{KT}(\hat{\pi}, V) - d_{KT}( \pi,V)$ whence the bound \eqref{e:guarantees}.  
\end{proof}

Let $B$ denote the right-hand side of the bound \eqref{e:guarantees} and let $L$ be any lower bound of the Kemeny score  $d_{KT}( \pi,V)$ such as the relaxed LP optimal value or  the  easy-to-compute bound 
\begin{align*} L_{min}^M &  = \sum_{\,\{x,y \}\subset C} \mathbf{1}((x,y),(y,x)\notin M) \min (a_{xy}, a_{yx}) \\
& + \mathbf{1}((x,y)\in M)a_{yx} + \mathbf{1}((y,x)\in M)a_{xy}.
\end{align*} 
\noindent 
Then we can deduce an upper bound the score gap of the approximate solution $\widehat{\pi}$ by 
$$
\frac{d_{KT}(\hat{\pi}, V) - d_{KT}( \pi,V)}{d_{KT}( \pi,V)} \leq 
\frac{B}{L}.$$

 \section{Local optimizer for feasible approximate solutions from $\Gamma$-XCC} 
 \label{s:local-opt}

For constrained weighted tournament,  M-ACP solutions frequently  face a serious problem. The relative ordering between the blocks in the initial partition of the M-XCC partition may cause violation to the initial constraints. M-ACP tends to lock in such violation which may result in a significant unresolved gap between M-ACP approximate solutions and true optimal solutions. More critically, M-ACP and $\Gamma$-XCC approximate solutions   may violate the immutable pairwise input constraints and thus they are infeasible for the constrained weighted tournament.  
\par 
A more pragmatic approach that solves the above issue is to start from the approximate solution $\pi^M$ produced by $\pmb\Gamma$-\textbf{XCC} with input pairwise constraints $M\subset C^2$. Here, the set of constraints $M$ is obtained from our method $\alpha$MOT* or $\alpha$MOTe* or G1 with/without the LowInf-Pairs booster. In particular, $M$ contains the set of initial immutable pairwise constraints $W$ of the problem. We then pick a random full ranking $\sigma^M$ satisfying the constraints $M$ that prioritizes as much as possible the relative ordering of elements according to the ranking $\pi^M$, for example using the Kahn algorithm. Finally we apply the local optimizer (Algorithm~\ref{algo:local-optimizer}) with half overlapping local windows alternatively several round. The resulting ranking denoted $\rho^M$ will serve as the approximate solution of the constrained weighted tournament. Note that $\rho^M$ is a feasible solution since each modification of $\sigma^M$ assures that there is no violation to the constraints $M$. 
At each iteration, the local optimization on disjoint segments of the ranking can be executed fully in parallel. Note that it suffices to compute the Kemeny score of $\rho^M$ then compare with the score of $\pi^M$, or to use the similar provable bound of Theorem~\ref{t:provable-guarantee}, to deduce a bound on the approximation factor for $\rho^M$.   \par 

\begin{algorithm}[hbt!]
\caption{Local optimizer}\label{algo:local-optimizer}
\DontPrintSemicolon 
\KwData{$C$,  $[\delta_{xy}]_{x,y \in C}$, constraints $M\subset C^2$ ranking $\sigma$, segment size $s$} 
\KwResult{ranking $\sigma^*$}
$n \gets |C|$; \, $\sigma^*\gets \sigma$;\,   \\ 
\For{$i = 0$ to $1$} 
{Decompose $\sigma^* = A L_1 \cdots L_k B$ into successive segments with $|A| = i \lfloor s/2 \rfloor$, $|L_1| = \cdots = |L_k| =  s $ where $k= \lfloor n  /s -i/2 \rfloor$ and $|B| = n - |A|-|L_1| -\cdots -|L_k|$;\\   
Replace each segment by a median of the induced constrained weighted tournament over the elements in that segment; \\
}
\KwRet{$\sigma^*$}
\end{algorithm}
More specifically, suppose that we are given a ranking $\sigma$  as an approximation of the medians of a  constrained weighted tournament over a set of candidates $C$. 
 In general, such an approximate solution $\sigma$ may not be {locally optimized}. More specifically, there may be a segment in $\sigma$ which is not a median of the induced election to elements of the segment. This observation motivates the post-processing application of the following local optimizer (Algorithm~\ref{algo:local-optimizer}).  Essentially, this local optimizer partitions the ranking $\sigma$ into successive segments of given equal size $s$ (except possibly the last segment) then rearranges elements within each segment if necessary to obtain a Kemeny optimized ordering over the same elements. The local optimizer repeats the same action but with a partition shifted to the right by $\lfloor s/2 \rfloor$ units. 
Since we optimize disjoint segments of the ranking, the whole process is fully parallelizable for each iteration. Algorithm~\ref{algo:local-optimizer} is the final piece for our  approximation of the medians and requires only a linear time complexity. 

\begin{theorem}
    The time complexity of Algorithm~\ref{algo:local-optimizer} is $O(n)$ for every fixed value of the user-specific parameter $s$ of the segment size.  
\end{theorem}

\begin{proof}
Let $t(s)$ be the worst case time complexity to compute a median of an election with $s$ candidates by the  method chosen in Algorithm~\ref{algo:local-optimizer}. Observe that Algorithm~\ref{algo:local-optimizer} performs $O(n)$ such computations. 
Therefore, the total time required for Algorithm~\ref{algo:local-optimizer}  is at most $O(t(s)n)$. 
\end{proof}

\section{The algorithm M-INCR-DEG}

Fix  a weighted tournament on a set of candidates $C$ with the weight matrix  $[a_{xy}]_{x,y \in C}$. Let $M\subset C^2$ be a set of constraints. To take into consideration the set $M$ in the algorithm  \textbf{INCR-DEG}, we define the matrix $[a_{xy}^M]_{x,y \in C}$ by 
\begin{equation}
a_{xy}^M 
= 
\begin{cases} 
1 & \text{ if } (x,y) \in M,\\
0 & \text{ if } (y,x) \in M,\\
a_{xy} & \text{ otherwise.}
\end{cases} 
\end{equation} 
As $a_{xy}  + a_{yx}=1$  and $M$ is anti-reflexive, $a_{xy}^M  + a_{yx}^M =1$ for all distinct $x,y$. 
Let $\pi^*$ be a median of the constrained weighted tournament such that $\pi^*$ satisfies $M$.  For  $x \in C$,  let $i_x^M = \sum_{y\in C} a_{xy}^M$ and 
$I_x^M = (n-1) - i_x^M$ and 

\begin{align*}
    L^{M-}_x & = \sum_{y \in C} a^M_{yx} \mathbf{1}( \pi^*(y)<\pi^*(x))  \\
    R^{M-}_x & = \sum_{y \in C} a^M_{yx}  \mathbf{1}( \pi^*(y)> \pi^*(x))  \\
    L^{M+}_x & = \sum_{y \in C}a^M_{xy}   \mathbf{1}( \pi^*(y)<\pi^*(x)). 
\end{align*}
We define \textbf{M-INCR-DEG} as the algorithm that takes as input the weight matrix   $[a_{xy}]_{x,y \in C}$ and the input constraints $M\subset C^2$ and returns a ranking denoted $\sigma^M$ that ranks the candidates $x$ of $C$ with decreasing index $i_x^M$. 
\begin{theorem}
\label{t:M-INCR-DEG-5-approx}
\textbf{\textup{M-INCR-DEG}} produces a 5-approximation of the
Kemeny problem for constrained weighted tournaments.
\end{theorem}
\noindent  
For the proof of  Theorem~\ref{t:M-INCR-DEG-5-approx}, using $a_{xy}^M  + a_{yx}^M =1$ for all distinct $x,y$, we have 
\begin{align}
\label{eq:proof-12-appro-2a}
    \pi^*(x) & =  L^{M-}_x  + L^{M+}_x \\
\label{eq:proof-12-appro-2b}
    I^M_x & = L^{M-}_x  + R^{M-}_x\\
    \label{eq:proof-12-appro-2z} 
     \sum_{x} \left( L_x^{M+} +R_x^{M-}\right)  & =   \sum_{x,y} \left( a^M_{yx}  \mathbf{1}( \pi^*(x) < \pi^*(y)) + a^M_{xy}  \mathbf{1}(\pi^*(y)<\pi^*(x)) \right)
    \end{align}
   \noindent
   Moreover, since $\pi^*$ satisfies $M$, we also have 
   \begin{align*}
    2 d_{KT}(\pi^*) & =  \sum_{x,y } \left( a_{yx}  \mathbf{1}( \pi^*(x) < \pi^*(y)) + a_{xy}  \mathbf{1}(\pi^*(y)<\pi^*(x)) \right)\\ 
& = \sum_{(x,y)\in M} \left( a_{yx}  \mathbf{1}( \pi^*(x) < \pi^*(y)) + a_{xy}  \mathbf{1}(\pi^*(y)<\pi^*(x)) \right)\\ 
& + \sum_{(y,x)\in M} \left( a_{yx}  \mathbf{1}( \pi^*(x) < \pi^*(y)) + a_{xy}  \mathbf{1}(\pi^*(y)<\pi^*(x)) \right)\\ 
& + \sum_{(x,y),(y,x)\notin M} \left( a_{yx}  \mathbf{1}( \pi^*(x) < \pi^*(y)) + a_{xy}  \mathbf{1}(\pi^*(y)<\pi^*(x)) \right)\\ 
& = \sum_{(x,y)\in M} a_{yx}  +  \sum_{(y,x)\in M} a_{xy}   \\ 
& + \sum_{(x,y),(y,x)\notin M} \left( a^M_{yx}  \mathbf{1}( \pi^*(x) < \pi^*(y)) + a^M_{xy}  \mathbf{1}(\pi^*(y)<\pi^*(x)) \right)\\
& = \sum_{(x,y)\in M} a_{yx}  +  \sum_{(y,x)\in M} a_{xy}     \\ 
& + \sum_{x,y} \left( a^M_{yx}  \mathbf{1}( \pi^*(x) < \pi^*(y)) + a^M_{xy}  \mathbf{1}(\pi^*(y)<\pi^*(x)) \right).
\end{align*} 
   \noindent
Combining with  \eqref{eq:proof-12-appro-2z} and  the triangle inequality, we deduce that 
   \begin{align} 
2 d_{KT}(\pi^*)  & = \sum_{x\in C} \left( L_x^{M+} +R_x^{M-}\right) +  \sum_{(x,y)\in M} a_{yx}  +  \sum_{(y,x)\in M} a_{xy} \nonumber \\
& \geq  \sum_{x\in C} \left( L_x^{M+} +R_x^{M-}\right)  \nonumber \\
\label{eq:proof-gamma-alpha-xcc-bound-ix}  
& \geq  \sum_{x\in C} | L_x^{M+}  - R_x^{M-}|  =  \sum_{x \in C}  |\pi^*(x)-I^M_x|  
\end{align} 
Finally, we deduce from \eqref{eq:proof-gamma-alpha-xcc-bound-ix},  Lemma~\ref{l:ordering-J-pi}, the triangle inequality, and Lemma~\ref{l:kendall-tau-vs-footnote} the following estimation  
\begin{align*}
4  d_{KT}(\pi^*) &  \geq  \sum_{x \in C}  |\pi^*(x)-I^M_x|  +  \sum_{x \in C}  |\pi^*(x)-I^M_x|  \\
& \geq  \sum_{x \in C}  |\pi^*(x)-I^M_x|  +  \sum_{x \in C}  |\sigma^M(x)-I^M_x|\\
& \geq \sum_{x \in C}  |\pi^*(x)-\sigma^M(x)|   = d_S(\pi^*, \sigma^M)\\ 
& \geq d_{KT} (\sigma^M) -d_{KT}(\pi^*), 
\end{align*}
and thus $d_{KT} (\sigma^M)  \leq 5 d_{KT}(\pi^*)$. This completes the proof of Theorem~\ref{t:M-INCR-DEG-5-approx}.

\section{Combining $\Gamma$-$\textbf{\textup{XCC}}$ and \textbf{\textup{M-INCR-INDEG}}}
Fix $p \in \mathbb{R}$. For every candidate $x$ in a constrained weighted tournament over $C$, we define 
\begin{equation}
p^M_x = p\gamma^M_x +(1-p) i^M_x  
\end{equation} 
Let  $D^M_x = I^M_x -  J^M_x$ 
where  $I^M_x = (n-1) - i^M_x$ and $J^M_x = (n-1) - \gamma^M_x$. 
Let $P^M_x= (n-1) - p^M_x= 
    P^M_x= I^M_x - pD^M_x$. 
Let $\bm \Gamma_p$\textbf{-XCC}  be the algorithm which returns a ranking $\pi_p^M$ by ordering the candidates  by their indexes $P^M_x$ in an increasing manner, where ties are broken arbitrarily. 

\begin{lemma}  
\label{l:combine-gamma-xcc-incre-indeg-1}
     We have $\displaystyle \sum_{x \in C} |\pi^*(x) - P_x| \leq {2(1+p)}d_{KT}(\pi^*)$. 
\end{lemma} 

\begin{proof} 
The proof is a direct generalization of the proof of Lemma~\ref{l:constant-approx-lemma-2-alpha} (and the relation~\ref{eq:proof-gamma-alpha-xcc-bound-ix}): it suffices to replace $J^M_x$ by $  P^M_x= I^M_x - pD^M_x$ and replace  $D^M_x$ by $|p|D^M_x$ in the estimations. 
\end{proof} 

\begin{theorem} 
 For every $p\in \mathbb{R}$,  the algorithm   $\bm \Gamma_p$\textbf{\textup{-XCC}} is a $(5+4|p|)$-approximation of the Kemeny problem for constrained weighted tournaments.  
\end{theorem} 

\begin{proof}
Let $\pi^*$ be a median of the constrained weighted tournament. By Lemma~\ref{l:combine-gamma-xcc-incre-indeg-1}, the triangle inequality, and similar statements of Lemma~\ref{l:ordering-J-pi} and Lemma~\ref{l:kendall-tau-vs-footnote} for $\pi^M_p$ and $P^M_x$ instead of $\pi^M$ and $J_x^M$,  we find that  
    \begin{align*}
 4(1+|p|) d_{KT}(\pi^*) & \geq \sum_{x \in C} |\pi^*(x) - P^M_x| + \sum_{x \in C} |\pi^*(x) - P^M_x| \\
    & \geq \sum_{x \in C} |\pi^*(x) - P^M_x| + \sum_{x \in C} |\pi^M_p(x) - P^M_x|\\
    & \geq \sum_{x \in C} |\pi^*(x) - \pi^M_p(x)| \\
    & \geq  d_{KT}(\pi^M_p) -   d_{KT}(\pi^*)
\end{align*}
whence the conclusion. 
\end{proof}
By varying $p\in [0,1]$, we obtain a family of high quality approximate rankings $\pi_p^M$ of the constrained weighted tournament. It suffices to choose the best solution among them. 
 
\section{Feasible solutions with bounded constant approximation factors of the Kemeny problem for constrained elections}
\label{s:election-properties-LP-based-solution}

In this section, we describe two polynomial time algorithms \textbf{\textup{{LP}-M-INCR-DEG}} and \textbf{\textup{LP}}-$\pmb {\Gamma}$-\textbf{\textup{XCC}} 
which base on LP solutions and produce feasible approximate solutions with bounded constant-approximation factors of the Kemeny aggregation ranking problem for  constrained elections. In particular, with the price of solving in polynomial time an LP program,  we solve the feasibility issue of the constant-factor approximation algorithms discussed in the previous sections. 
\par 
Suppose that $M\subseteq C^2$ is a set of immutable input pairwise constraints of a weighted tournament $[a_{xy}]_{x,y \in C}$ over $C$. 
We identify each ranking $\pi$ by a binary vector $[c_{xy}]_{x,y\in C}$ where $c_{xx}=0$ for all $x\in C$, $c_{xy}=1$ if $x$ is ranked before $y$ in $\pi$, and $c_{yx}=1$ otherwise. 
Consequently, a feasible $\pi$ of the constrained weighted tournament must satisfy $c_{x,y}=1$ for all $(x,y)\in M$. 
By incorporating the probability constraints $c_{xy} + c_{yx} = 1$ and the transitivity constraints $c_{xy} + c_{yz} + c_{zx} \geq 1$, we obtain the standard LP relaxation of the constrained Kemeny aggregation problem: 
\begin{align}
\label{e:IP}
  \min \quad & \sum_{\{x,y\}  \in C}  c_{xy} a_{yx} + c_{yx} a_{xy} \\
\mathrm{s.t.}\quad  &  c_{xy} \in [0,1], \, c_{xy} + c_{yx} = 1,   \quad  \forall  \{x,y\}  \in C \nonumber  \\ 
& c_{xy} + c_{yz} + c_{zx} \geq 1, \quad \, \forall  \{x,y,z\}  \in C \nonumber 
\\
& c_{xy}  = 1,  \quad  \forall  (x,y) \in M \nonumber  
\end{align}
\noindent 
Let $[c^*_{xy}]_{x,y\in C}$ be an optimal LP solution of the above LP model.  We define \textbf{\textup{LP}}-$\pmb {\Gamma}$-\textbf{\textup{XCC}}  as the algorithm that takes as input the weight matrix $[a_{xy}]_{x,y \in C}$, the input constraints $M\subset C^2$, and returns a ranking denoted $\pi_{LP}^M$ that ranks the candidates $x$ of $C$ with decreasing index $\gamma_{x,LP}^M$  (with ties broken arbitrarily) where 
\begin{equation}
\label{eq:gamma-x-lp-m} 
\gamma_{x,LP}^M = \sum_{y \in C \setminus\{x\}} \langle c^*_{xy} \rangle
\end{equation} 
\noindent
where   $\langle b\rangle = 0$ if $0\leq b\leq \frac{1}{2}$ and  $\langle b  \rangle = 1$ if $\frac{1}{2}<b\leq 1$. 
Similarly, let \textbf{\textup{{LP}-M-INCR-DEG}} be the algorithm that takes as input the weight matrix $[a_{xy}]_{x,y \in C}$, the input constraints $M\subset C^2$, and returns a ranking denoted $\sigma_{LP}^M$ that ranks the candidates $x$ of $C$ with decreasing index $i_{x,LP}^M$ (with ties broken arbitrarily) where 
\begin{equation}
\label{eq:i-x-lp-m} 
i_{x,LP}^M = \sum_{y \in C\setminus\{x\}}  c^*_{xy}. 
\end{equation}

Note that computing the optimal LP solution $[c^*_{xy}]_{x,y\in C}$ as well as computing and sorting the indexes $i_{x,LP}^M$ and $\gamma_{x,LP}^M$ 
  all have polynomial time complexity. It follows that both algorithms \textbf{\textup{{LP}-M-INCR-DEG}} and \textbf{\textup{LP}}-$\pmb {\Gamma}$-\textbf{\textup{XCC}} have polynomial time complexity. 
Moreover, the lemma below shows that the solutions $\pi_{LP}^M$ and  $\sigma_{LP}^M$ are all feasible solutions of the constrained Kemeny aggregation ranking problem.  

\begin{lemma}
The rankings $\pi_{LP}^M$ and  $\sigma_{LP}^M$ are feasible solutions of the constrained Kemeny aggregation ranking problem, i.e., $\pi_{LP}^M(x) < \pi_{LP}^M(y)$ and $\sigma_{LP}^M(x) < \sigma_{LP}^M(y) $ for all $(x,y)\in M$.   
\end{lemma} 

\begin{proof}
Let $(x,y)\in M$. Then by the formulation \eqref{e:IP} of the LP model, we must have $c^*_{xy}  = 1$ and thus $c^*_{yx} = 1- c^*_{xy} =0$ by the probability constraint. 
Let $z\in C\setminus\{x,y\}$. The transitivity constraint for the triple $(y,x,z)$ implies that $c^*_{yx} + c^*_{xz} + c^*_{zy} \geq 1$.  
It follows that $ c^*_{xz} + c^*_{zy} \geq 1$ and thus  $c^*_{xz} \geq  1- c^*_{zy} =  c^*_{yz}$. In particular, we also have $\langle c^*_{xz} \rangle \geq \langle c^*_{yz} \rangle$ for all $z\in C\setminus\{x,y\}$ since the function $ b\mapsto \langle b \rangle$ is nondecreasing. We can now estimate  
\begin{align*}
i_{x,LP}^M  & = c^*_{xy}  +  \sum_{z \in C \setminus\{x,y\}}  c^*_{xz} 
 \geq 1+  \sum_{z \in C \setminus\{x,y\}}   c^*_{yz}   
 = 1 + i_{y,LP}^M
> i_{y,LP}^M,\\
\gamma_{x,LP}^M  & =  \langle c^*_{xy}  \rangle +  \sum_{z \in C \setminus\{x,y\}}  \langle c^*_{xz} \rangle
 \geq 1+  \sum_{z \in C \setminus\{x,y\}}  \langle c^*_{yz}   \rangle
 = 1 + \gamma_{y,LP}^M
> \gamma_{y,LP}^M.
\end{align*}
Therefore, $x$ must be ranked before $y$ in the rankings $\pi_{LP}^M$ and $\sigma_{LP}^M$ for all $(x,y)\in M$. 
We conclude that $\pi_{LP}^M$ and $\sigma_{LP}^M$ are feasible solutions of the  constrained Kemeny aggregation ranking problem. 
\end{proof}

The next two theorems provide a constant bound for the approximation factors of  \textbf{\textup{{LP}-M-INCR-DEG}}  and  \textbf{\textup{LP}}-$\pmb {\Gamma}$-\textbf{\textup{XCC}} in the case of constrained elections.

\begin{theorem}
\label{t:4-appr-LP-m-incr-deg}
   The algorithm  \textbf{\textup{{LP}-M-INCR-DEG}}   is a 4-approximation of  the Kemeny problem for constrained elections. 
\end{theorem}

\begin{proof}
Let $\sigma_1, \dots, \sigma_m$ be the collection of all votes (repetition allowed) in an election over $C$ with (a posteriori) immutable input  pairwise constraints $(x,y)\in M$. 
Let $\sigma_{LP}^M$ be the solution returned by \textbf{\textup{{LP}-M-INCR-DEG}}  with input constraints $M$ and input weight matrix $[a_{xy}]_{x,y \in C}$ induced by the above votes $\sigma_1, \dots, \sigma_m$. 
Let $\pi^*$ be a median of the constrained election and let  $[c^*_{xy}]_{x,y\in C}$ be an optimal solution of the LP model \eqref{e:IP}. 
We define $u_i=[u_i(x,y)]_{x,y \in C}$  where    $u_i(x,y)= \mathbf{1}(\sigma_i(x) < \sigma_i(y))$.   
We set $ I_{x,LP}^M  = (n-1) -  i_{x,LP}^M$
 and  $c^*_{xx} = u_i(x,x)=0$ for all $x \in C$. Since the optimal objective value of the LP model \eqref{e:IP}  is a lower bound of $d_{KT}(\pi^*)$, we have 
\begin{align}
\label{eq:t:4-appr-LP-m-incr-deg-1}
 d_{KT}(\pi^*)  \geq  \sum_{\{x,y\}  \subset C}  c^*_{xy} a_{yx} + c^*_{yx} a_{xy}. 
 \end{align} 
 Moreover,  observe that by the probability constraint $c^*_{xy} + c^*_{yx}=1$, we have 
 \begin{align}
\label{eq:t:4-appr-LP-m-incr-deg-2}
 4 m  \sum_{\{x,y\}  \subset C}  c^*_{xy} a_{yx} + c^*_{yx} a_{xy}  & =  2\sum_{(x,y)  \in C^2}  c^*_{xy} ma_{yx} + c^*_{yx} ma_{xy}  \nonumber \\ 
& =  2\sum_{(x,y)  \in C^2}   \sum_{i =1}^m c^*_{xy}  u_i(y,x) +  c^*_{yx} u_i(x,y) \nonumber \\
& = 2 \sum_{(x,y)  \in C^2}    \sum_{i =1}^m |c^*_{xy} -u_i(x,y) |.
\end{align} 
By the triangle inequality, we can estimate \begin{align} 
\label{eq:t:4-appr-LP-m-incr-deg-3}
 2 \sum_{(x,y)  \in C^2}    \sum_{i =1}^m |c^*_{xy} -u_i(x,y) | 
  &  \geq   2  \sum_{i =1}^m \sum_{x  \in C}  |  \sum_{y \in C}   c^*_{xy} -u_i(x,y) |  \nonumber \\
  & =  2\sum_{i =1}^m \sum_{x  \in C}  |  \sum_{y \in C}   c^*_{xy} - \sum_{y \in C} u_i(x,y) | \nonumber \\
     & =  2\sum_{i =1}^m \sum_{x  \in C}  |     i_{x,LP}^M  - (n-1) + \sigma_i(x)| \nonumber \\ 
          & = 2 \sum_{i =1}^m \sum_{x  \in C}  |  \sigma_i(x) -  I_{x,LP}^M |. 
\end{align}
By Lemma~\ref{l:kemeny-spearman} and Lemma~\ref{l:ordering-J-pi}, we obtain 

\begin{align}
\label{eq:t:4-appr-LP-m-incr-deg-4}
   2 \sum_{i =1}^m \sum_{x  \in C}  |  \sigma_i(x) -  I_{y,LP}^M | 
                 & \geq  \sum_{i =1}^m \sum_{x  \in C} \left(  |  \sigma_i(x) -  I_{y,LP}^M | +  |  \sigma_{LP}^M(x) -  I_{y,LP}^M |  \right) \nonumber \\
          & \geq  \sum_{i =1}^m \sum_{x  \in C}  |  \sigma_i(x) -  \sigma_{LP}^M(x) |  = \sum_{i =1}^m  d_S(\sigma_{LP}^M,\sigma_i) \nonumber \\
          & \geq \sum_{i =1}^m  d_{KT}(\sigma_{LP}^M,\sigma_i) = m d_{KT}(\sigma_{LP}^M).  
\end{align}
\noindent 
Combining the estimations \eqref{eq:t:4-appr-LP-m-incr-deg-1}, \eqref{eq:t:4-appr-LP-m-incr-deg-2}, \eqref{eq:t:4-appr-LP-m-incr-deg-3} and \eqref{eq:t:4-appr-LP-m-incr-deg-4}, it follows that $4md_{KT}(\pi^*) \geq md_{KT}(\sigma_{LP}^M)$ and thus $4d_{KT}(\pi^*) \geq d_{KT}(\sigma_{LP}^M)$. We conclude that \textbf{\textup{{LP}-M-INCR-DEG}}   is a 4-approximation of  the Kemeny problem for constrained elections. 
\end{proof}

\noindent We have a similar result for   \textbf{\textup{LP}}-$\pmb {\Gamma}$-\textbf{\textup{XCC}}  albeit with a looser bound. 

  \begin{theorem}
\label{t:4-appr-LP-gamma-alpha-xcc}
   The algorithm   \textbf{\textup{LP}}-$\pmb {\Gamma}$-\textbf{\textup{XCC}}  is an 8-approximation of  the Kemeny problem for constrained elections. 
\end{theorem} 

\begin{proof} 
We keep the notations and definitions as in the above proof of Theorem~\ref{t:4-appr-LP-m-incr-deg}. 
Let $\pi_{LP}^M$ be the solution returned by \textbf{\textup{LP}}-$\pmb {\Gamma}$-\textbf{\textup{XCC}}  with input constraints $M$ and input weight matrix $[a_{xy}]_{x,y \in C}$. Let  $J_{x,LP}^M  = (n-1) -  \gamma_{x,LP}^M$. Replacing $\sigma_{LP}^M$, $i_{x,LP}^M$, $I_{x,LP}^M$, $c^*_{xy}$ by 
$\pi_{LP}^M$, $j_{x,LP}^M$, $J_{x,LP}^M$, and $\langle c^*_{xy} \rangle$ respectively in the relations \eqref{eq:t:4-appr-LP-m-incr-deg-2}, \eqref{eq:t:4-appr-LP-m-incr-deg-3} and \eqref{eq:t:4-appr-LP-m-incr-deg-4}, we obtain by the same proof of these estimates that 
\begin{align} 
\label{eq:t:4-appr-LP-m-incr-deg-5}
4 \sum_{\{x,y\}  \subset C}  \langle c^*_{xy} \rangle a_{yx} + \langle c^*_{yx}  \rangle a_{xy}   \geq  d_{KT} (\pi_{LP}^M).
\end{align}  
\noindent 
Since $ \langle c^*_{xy}\rangle \leq 2 c^*_{xy}$ for all $x\neq y$, it follows that 
\begin{align} 
\label{eq:t:4-appr-LP-m-incr-deg-6}
 2 d_{KT}(\pi^*)  & \geq  2  \sum_{\{x,y\}  \subset C}    c^*_{xy}  a_{yx} +  c^*_{yx}  a_{xy}  \nonumber   \\
& = \sum_{\{x,y\}  \subset C}    2 c^*_{xy}      a_{yx} + 2 c^*_{yx}  a_{xy} \nonumber   \\
& \geq \sum_{\{x,y\}  \subset C}  \langle c^*_{xy} \rangle a_{yx} + \langle c^*_{yx}  \rangle a_{xy}. 
\end{align} 
\noindent 
Combining the relations \eqref{eq:t:4-appr-LP-m-incr-deg-5}, and  \eqref{eq:t:4-appr-LP-m-incr-deg-6}, we obtain $d_{KT} (\pi_{LP}^M) \leq 8 d_{KT} (\pi^*)$ and the conclusion follows. 
\end{proof}

\section{Discussion}
 \label{s:discussion}
  
 In this paper, we address the constrained Kemeny aggregation ranking problem with a posteriori immutable  input pairwise constraints. We present  various algorithms $\bm \Gamma_p$\textbf{-XCC}, \textbf{M-ACP}, \textbf{M-INCR-DEG}, $\bm \Gamma_p$\textbf{-XCC},  \textbf{\textup{{LP}-M-INCR-DEG}}, and  \textbf{\textup{LP}}-$\pmb {\Gamma}$-\textbf{\textup{XCC}} to produce robust and scalable approximate solutions with constant-approximation factors and polynomial runtimes.  
The algorithm  $\bm \Gamma$\textbf{-XCC} stands out as it satisfies the constrained Condorcet consistency property while   
 \textbf{\textup{{LP}-M-INCR-DEG}} and  \textbf{\textup{LP}}-$\pmb {\Gamma}$-\textbf{\textup{XCC}} solve the feasibility issue. To obtain the best theoretical features of these algorithms, it suffices to  apply  \textbf{\textup{{LP}-M-INCR-DEG}} or  \textbf{\textup{LP}}-$\pmb {\Gamma}$-\textbf{\textup{XCC}} on each block in the $M$-XCC partition of the constrained Kemeny problem. The resulting rankings satisfy the constrained Condorcet consistency property, the feasibility condition, and the bounded constant-approximation  factors. 
  \par 
   We also describe  methods $\alpha$MOT*, $\alpha$MOTe*, G1 to produce exact valid pairwise constraints and approximate one in order to facilitate the search for both approximate and exact solutions. As additional remarks, we observe from Figure~\ref{fig:correct-instances}  that the greedy method G1  excels and can even provide valid constraints when they are not too greedy, that is, when the number of pairs solved by them is not too close to the maximal number of pairs.  
 When they become too greedy, some of the constraints found by G1 may lead to suboptimal solutions.
  We note that higher approximation thresholds  $h$ do not always guarantee better results.  
 For approximate methods \textbf{M-ACP}, we recommend running in parallel ACP, $\alpha$ACP, $\alpha$ACPe,  ACP-G1 with varying approximation threshold $h$, choose the output ranking with the lowest Kemeny score and compare with the approximate solutions returned by , $\bm \Gamma$\textbf{-XCC} and  \textbf{M-INCR-DEG}, and their combinations $\bm \Gamma_p$\textbf{-XCC}.

 \begingroup
\setstretch{1.0}  \footnotesize    
\bibliographystyle{elsarticle-harv} 
\bibliography{Kemeny}    
\endgroup

 \end{document}